\documentclass[preprint,11pt]{elsarticle}

\usepackage{amssymb}
\usepackage{graphicx}
\usepackage{latexsym}
\usepackage{bm}
\usepackage{amsmath}
\journal{Progress in Particle and Nuclear Physics}

\begin{document}
\newcommand{\leftg}{\langle \phi_0 |}
\newcommand{\rightg}{| \phi_0 \rangle}
\newcommand{\vect}[1]{\stackrel{\rightarrow}{#1}}
\newcommand{\rvec}{\vec{r}}
\newcommand{\nn}{\nonumber}
\newcommand{\sigmavec}{\boldsymbol{\mathbf\sigma}}
\newcommand{\tauvec}{\boldsymbol{\mathbf\tau}}
\def\rbot{r_{\scriptscriptstyle{\bot}}}
\def\half{{{\scriptstyle 1}\over{\scriptstyle 2}}}
\def\halfx{{\scriptstyle {1\over 2}}}
\newcommand{\bff}[1]{{\mbox{\boldmath $#1$}}}
\newcommand{\D}{\displaystyle}
\newcommand{\lp}{\left}
\newcommand{\rp}{\right}
\newcommand{\fud}[2]{\frac{\delta #1}{\delta #2}}
\newcommand{\pad}[2]{\frac{\partial #1}{\partial #2}}
\newcommand{\bra}[1]{\left\langle #1 \right|}
\newcommand{\ket}[1]{\left| #1 \right\rangle}
\newcommand{\vev}[2]{\bra{#1} #2 \ket{#1}}
\newcommand{\boma}[1]{\mbox{\boldmath $#1$}}
\newcommand{\identity}{{\sf 1 \hspace{-0.15em}
                        \rule{0.087em}{1.5ex}
                        \rule{0.12em}{0.1ex}
                        \hspace{-0.3em}
                        \rule[1.4ex]{0.12em}{0.1ex}
                        \hspace{0.3em}}}

\begin{frontmatter}



\title{Relativistic Nuclear Energy Density Functionals: Mean-Field and Beyond}


\author{T. Nik\v si\' c and D. Vretenar}

\address{Physics Department, Faculty of Science, University of Zagreb,
10000 Zagreb, Croatia}

\author{P. Ring}
\address{Physik-Department der Technischen Universit\"at M\"unchen,
D-85748 Garching, Germany}

\begin{abstract}
Relativistic energy density functionals (EDF) have become a standard tool for
nuclear structure calculations, providing a complete and accurate, global
description of nuclear ground states and collective excitations.
Guided by the medium dependence of the
microscopic nucleon self-energies in nuclear matter, semi-empirical
functionals have been adjusted to the nuclear matter equation of state and
to bulk properties of finite nuclei, and applied to studies of
arbitrarily heavy nuclei, exotic nuclei far from stability, and even
systems at the nucleon drip-lines. REDF-based structure models
have also been developed that go beyond the static
mean-field approximation, and include collective correlations related to the
restoration of broken symmetries and to fluctuations of collective variables.
These models are employed in analyses of structure phenomena related to
shell evolution, including detailed predictions of excitation spectra and
electromagnetic transition rates.
\end{abstract}


\end{frontmatter}

\bigskip
\section{Introduction}
\label{Intro}

Nuclear energy density functionals (EDF) presently provide
the most complete and accurate description of ground-state properties
and collective excitations over the whole nuclide chart.
Among the microscopic approaches to the nuclear many-body
problem, probably no other method achieves comparable
global accuracy at the same computational cost, and it is the
only one that can describe the evolution of structure phenomena
from relatively light systems to superheavy nuclei, and from the
valley of $\beta$-stability to the particle drip-lines \cite{LNP.641,BHR.03}.

In practical implementations the EDF framework is realized on two
specific levels. The basic implementation is in terms of
self-consistent mean-field (SCMF) models, in which an EDF is constructed
as a functional of one-body nucleon density matrices that correspond to
a single product state -- Slater determinant of single-particle or
single-quasiparticle states. The SCMF approach to nuclear structure is analogous
to Kohn-Sham density functional theory (DFT) \cite{KS.65,Kohn.99}. DFT enables
a description of quantum many-body systems in terms of a universal energy
density functional. Universal in the sense that, for a given inter-particle interaction,
it has the same functional form for all systems. Nuclear SCMF models
effectively map the many-body problem onto a one-body
problem, and the exact EDF is approximated by simple, mostly analytical,  
functionals of powers and gradients of ground-state nucleon densities and currents,
representing distributions of matter, spins, momentum and kinetic
energy. In principle the nuclear EDF can incorporate short-range
correlations related to the repulsive core of the inter-nucleon
interaction, and long-range correlations mediated by nuclear
resonance modes. Even though it originates in the effective interaction
between nucleons, a generic density functional is not necessarily related
to any given nucleon-nucleon (NN) potential and, in fact, some of the most successful
modern functionals are entirely empirical. Of course it would be desirable to have a fully
microscopic foundation for a universal density functional, and this is certainly one of the major
challenges for the framework of nuclear EDFs. Because it includes
correlations, the self-consistent Kohn-Sham approach goes beyond the
Hartree-Fock approximation, and it also has the advantage of being a local
scheme. Its usefulness, however, crucially depends on our ability to
construct accurate approximations for the most important part of the
functional, that is, the universal exchange-correlation functional \cite{DG.90}.
In an ``ab initio" approach one might start from a Hamiltonian that describes
two-nucleon and few-body scattering and bound-state observables \cite{DFP.10},
or an effective field theory of low-energy in-medium NN interactions
can be used to build approximations to the exact exchange-correlation
functional \cite{Vre.08}. However, even if a fully microscopic EDF is eventually developed,
the parameters of that functional will still have to be fine tuned to structure data of finite
nuclei. This is because data on nucleon-nucleon scattering and few-nucleon systems,
or gross properties of infinite nuclear matter, cannot determine the density functional
to a level of accuracy necessary for a quantitative description of medium-heavy and
heavy nuclei.

When considering applications, however, an important challenge for the
framework of EDF is the systematic treatment of collective correlations related to
restoration of broken symmetries and fluctuations in collective coordinates.
A static nuclear EDF is of course characterized by symmetry
breaking -- translational, rotational, particle number, and can only
provide an approximate description of bulk ground-state properties.
To calculate excitation spectra and electromagnetic transition rates
in individual nuclei, it is necessary to extend the Kohn-Sham EDF
framework, that is the SCMF scheme, to include correlations that
arise from symmetry restoration and fluctuations around the mean-field
minimum. Collective correlations are sensitive to shell effects, display
pronounced variations with particle number and, therefore, cannot be
incorporated in a universal EDF. On the second level that takes into
account collective correlations through the restoration of broken symmetries
and configuration mixing of symmetry-breaking product states, the
many-body energy takes the form of a functional of all transition
density matrices that can be constructed from the chosen set of
product states. This set is chosen to restore symmetries or/and to
perform a mixing of configurations that correspond to specific collective
modes using, for instance, the (quasiparticle) random-phase approximation
(QRPA) or the Generator Coordinate Method (GCM).
The latter includes correlations related to finite-size fluctuations in a collective
degree of freedom, and can be also used to restore selection rules that are
crucial for spectroscopic observables.

An important class of nuclear structure models belongs to the framework of
relativistic energy density functionals. In particular, a number of very successful
relativistic mean-field (RMF) models have been constructed based on the
framework of quantum hadrodynamics (QHD) \cite{SW.86,SW.97}.
There are important advantages in using functionals with manifest
covariance \cite{FS.00}. The most obvious is the natural inclusion of the
nucleon spin degree of freedom, and the resulting nuclear spin-orbit potential
which emerges automatically with the empirical strength in a covariant formulation.
The consistent treatment of large, isoscalar, Lorentz scalar and vector self-energies
provides a unique parametrization of time-odd components of the nuclear
mean-field, i.e. nucleon currents, which is absent in the non-relativistic
representation of the energy density functional. The empirical pseudospin
symmetry in nuclear spectroscopy finds a natural explanation in terms of
relativistic mean fields \cite{Joe.05}. On a microscopic level, it has been
argued \cite{FS.00} that a covariant formulation of nuclear dynamics
manifests the true energy scales of QCD in nuclei, and is consistent with
the nonlinear realization of chiral symmetry through the implicit inclusion of
pion-nucleon dynamics in the effective nucleon self-energies. A covariant
treatment of nuclear matter provides a distinction between scalar and
four-vector nucleon self energies, leading to a very natural saturation
mechanism.

RMF-based models have been very successfully employed in
analyses of a variety of nuclear structure phenomena,
not only in nuclei along the valley of $\beta$-stability,
but also in exotic nuclei with extreme isospin
values and close to the particle drip lines. Applications
have reached a level of sophistication and
accuracy comparable to the non-relativistic
Hartree-Fock-Bogoliubov approach based on Skyrme
functionals or Gogny effective
interactions \cite{BHR.03,VALR.05,Meng.06,PVKC.07}.
Relativistic EDFs have mostly been applied in the description of
ground-state properties and excitation energies of giant resonance
at the self-consistent mean-field level, taking into account pairing
correlations in open-shell nuclei in the  Hartree-(Fock)-Bogoliubov
framework, and performing consistent (Q)RPA calculations of
small-amplitude collective motion. However, for relativistic
structure models to make detailed spectroscopic predictions,
symmetries broken by the static nuclear mean field must be restored, and
fluctuations around the mean-field minimum must be taken into account.
While symmetry restoration and configuration mixing calculations have
routinely been applied with non-relativistic density functionals since
many years, it is only more recently that this type of structure models
have been developed using relativistic density
functionals \cite{NVR.06a,NVR.06b,Ni.09,Yao.09,Yao.10}.

In this work we review recent advances in the framework of
relativistic EDFs and, in particular, the latest extensions that include the
treatment of collective correlations. Section \ref{REDF} introduces the
general framework of REDFs. In Section \ref{DD-PC1} we review
a class of semi-empirical functionals and, in particular, the functional
DD-PC1 that will be used in illustrative calculations throughout this work.
For a quantitative analysis of open-shell nuclei it is necessary to consider
pairing correlations, and in Section \ref{RHB} we introduce the
relativistic Hartree-Bogoliubov model for triaxial nuclei, with a separable
pairing interaction. The treatment of collective correlations is reviewed and
illustrated with a number of examples in Section \ref{GCM} (symmetry
restoration and configuration mixing calculations) and in Section \ref{H_coll}
(collective Hamiltonian in five dimensions). Section \ref{Summary} summarizes the
results and ends with an outlook for future studies.
\section{Relativistic Energy Density Functionals}
\label{REDF}
\subsection{Quantum Hadrodynamics}

In conventional quantum hadrodynamics (QHD)  \cite{SW.86,SW.97}
a nucleus is described as a system of Dirac nucleons coupled to exchange
mesons through an effective Lagrangian. The isoscalar scalar $\sigma$ meson,
the isoscalar vector $\omega$ meson, and the isovector vector $\rho$ meson
build the minimal set of meson fields that, together with the electromagnetic
field, is necessary for a description of bulk and single-particle nuclear
properties. In the {\em mean-field} approximation the meson-field operators
are replaced by their expectation values in the nuclear ground state.
In addition, a quantitative treatment of nuclear matter and finite
nuclei necessitates a medium dependence of effective mean-field interactions
that takes into account higher-order many-body effects. A medium dependence
can either be introduced by including nonlinear meson self-interaction terms
in the Lagrangian, or by assuming an explicit density dependence for the
meson-nucleon couplings. The former approach has been adopted in the
construction of several successful phenomenological RMF interactions, for
instance, the very popular NL3~\cite{NL3}, or the more recent PK1, PK1R
\cite{PK.04} and FSUGold \cite{FSUGold} parametrizations of the effective
Lagrangian. In the latter case, the density dependence of the meson-nucleon
vertex functions can be parameterized starting from microscopic Dirac-Brueckner
calculations of symmetric and asymmetric nuclear matter
\cite{FLW.95,JL.98,HKL.01} or it can be fully phenomenological
\cite{TW.99,NVFR.02,LNVR.05}, with parameters adjusted to data on finite
nuclei and empirical properties of symmetric and asymmetric nuclear matter.

At the energy scale characteristic for nuclear binding and low-lying excited
states, meson exchange ($\sigma$, $\omega$, $\rho$, $\ldots$) is just a
convenient representation of the effective nuclear interaction. The exchange
of heavy mesons is associated with short-distance dynamics that cannot be
resolved at low energies, and therefore in each channel (scalar-isoscalar,
vector-isoscalar, scalar-isovector, and vector-isovector) meson exchange can be
replaced by the corresponding local four-point (contact) interactions between
nucleons. The self-consistent relativistic mean-field framework can be
formulated in terms of point-coupling nucleon interactions. When applied in
the description of finite nuclei, relativistic mean-field point-coupling
(RMF-PC) models \cite{MNH.92,Hoch.94,FML.96,RF.97,BMM.02} produce results that
are equivalent to those obtained in the meson exchange picture. Of course,
also in the case of contact interactions, medium effects can be taken into
account by the inclusion of higher-order interaction terms, for instance,
six-nucleon vertices $(\bar{\psi}\psi)^{3}$, and eight-nucleon vertices
$(\bar{\psi}\psi)^{4}$ and $[(\bar{\psi}\gamma_{\mu}\psi)(\bar{\psi}%
\gamma^{\mu}\psi)]^{2}$, or it can be encoded in the effective couplings, i.e.
in the density dependence of strength parameters of the interaction in the
isoscalar and isovector channels. Although a number of point-coupling models
have been developed over the years, it is only more recently that
phenomenological parametrizations have been adjusted and applied in the
description of finite nuclei on a level of accuracy comparable to that of
standard meson-exchange effective interactions \cite{BMM.02,NVR.08}.

The relation between the two representations: finite-range
(meson exchange) and zero-range (point-coupling), is straightforward in nuclear
matter because of constant nucleon scalar and vector densities.
The Klein-Gordon equations of the
meson-exchange model with meson masses $m_{\phi}$ and
density-dependent couplings $g_{\phi}(\rho)$, are replaced by the corresponding
point-coupling interaction terms with strength parameters
$g_{\phi}^{2}/m_{\phi}^{2}$. In finite nuclei, however, the problem is not so
simple. Because of the radial dependence of the densities, the expansion of the
meson propagator in terms of $1/m_{\phi}^{2}$ leads to an infinite series of
gradient terms. In practice this series has to be replaced by a finite number
of terms with additional phenomenological parameters
adjusted to low-energy data. A number of studies have shown that, both for
finite-range and for point-coupling mean-field models, the empirical data set
of ground-state properties of finite nuclei can determine only a relatively
small number of parameters in the general expansion of the effective Lagrangian
in powers of the fields and their derivatives. It is therefore not a priori
clear how to select the set of point-coupling interaction terms that will
describe structure properties at the same level of accuracy as the
meson-exchange models. The mapping of a phenomenological
finite-range interaction with density-dependent meson-nucleon
couplings (DD-ME2) on the zero-range (point-coupling) relativistic mean-field
framework was considered in Ref.~\cite{NVLR.08}.
A family of point-coupling effective interactions was constructed with different
values of the strength parameter of the isoscalar-scalar derivative term.
In the meson-exchange picture this corresponds to different values of the
$\sigma$-meson mass. The parameters of the isoscalar-scalar and
isovector-vector channels of the point-coupling interactions were adjusted to
nuclear matter and ground-state properties of finite nuclei. By comparing
results for infinite and semi-infinite nuclear matter, ground-state masses,
charge radii, and collective excitations, constraints were placed on the
parameters of phenomenological point-coupling relativistic effective interaction.

\subsection{Elements of Density Functional Theory}

Density Functional Theory (DFT) is one of the most popular and successful 
``ab initio" approaches to the structure of quantum many-body systems (atoms, molecules, solids).  
Probably no other method achieves comparable 
accuracy at the same computational cost. The basic concept is that the ground-state
properties of a stationary many-body system can be represented in terms 
of the ground-state density alone. Since the density $\rho(\bf{r})$ is a function of 
only three spatial coordinates, rather than the 3N coordinates of the N-body 
wave function, DFT is computationally feasible even for large systems. 

Most practical applications of Density Functional Theory  use the effective single-particle
Kohn-Sham (KS) equations  \cite{KS.65,Kohn.99}, introduced for an auxiliary system of N
non-interacting particles. According to the Hohenberg-Kohn theorem \cite{HK.64},
there exists a unique energy functional
\begin{equation}
E_s[\rho] = T_s[\rho] + \int d^3r~v_s({\bf r}) \rho({\bf r}) \; ,
\end{equation}
for which the variational equation yields the exact ground-state density
$\rho_s({\bf r})$. $T_s[\rho]$ is the universal kinetic energy functional of the
non-interacting system. The KS scheme is based on the following assertion:
for any interacting system, there exists a unique local
single-particle potential $v_s({\bf r})$, such that the exact ground-state density of the
interacting system equals the ground-state density of the auxiliary non-interacting system:
\begin{equation}
\rho({\bf r})=\rho_s({\bf r}) = \sum_i^{N} |\phi_i ({\bf r})|^2 \; ,
\label{occ}
\end{equation}
expressed in terms of the N lowest occupied single-particle orbitals -- solutions of the
Kohn-Sham equations:
\begin{equation}
\left [ - {{\nabla^2}/{2m}} + v_s({\bf r}) \right ] \phi_i ({\bf r})
= \varepsilon_i \phi_i ({\bf r}) \; .
\label{KS_eq}
\end{equation}
The uniqueness of $v_s({\bf r})$ follows from the Hohenberg-Kohn theorem
and the single-particle orbitals are unique functionals of the density:
$\phi_i ({\bf r}) = \phi_i ([\rho];{\bf r})$.

For a self-bound system like the atomic nucleus, the energy functional can be
decomposed into three separate terms:
\begin{equation}
\label{KS}
F[\rho] = T_{s}[\rho] + E_{H}[\rho] + E_{xc}[\rho] \; ,
\end{equation}
where $T_{s}$ is the kinetic energy of
the non-interacting A-nucleon system, $E_{H}$ is a Hartree
energy, and $E_{xc}$  denotes the exchange-correlation
energy which, by definition, contains everything else -- all the
many-body effects. The corresponding local {exchange-correlation potential}
is defined by:
\begin{equation}
v_{xc}[\rho]({\bf r})  = {{\delta E_{xc}[\rho]}\over {\delta \rho({\bf r})}} \; ,
\end{equation}
and thus
\begin{equation}
v_{s}[\rho]({\bf r}) =  v_{H}[\rho]({\bf r}) + v_{xc}[\rho]({\bf r})\; .
\label{v_KS}
\end{equation}
Since the effective potential depends on the ground-state density, the system of equations
(\ref{occ}), (\ref{KS_eq}), and (\ref{v_KS}) has to be solved self-consistently.
This is the Kohn-Sham scheme of density functional theory \cite{KS.65}.
By including correlation effects the KS framework goes beyond the
Hartree-Fock approximation but, in addition, it has the advantage of being
a {\em local} scheme. It is clear, however, that
the usefulness of the Kohn-Sham scheme crucially depends
on our ability to construct accurate approximations to the
exact exchange-correlation energy.
The true exchange-correlation energy functional is {\em universal},
i.e. given the inter-particle interaction, it has the same functional form for
all systems. One possible approach is to develop $E_{xc}$ from first principles
by incorporating known exact constraints. Another is empirical,
a parametric ansatz is optimized by adjusting it to a set of data.

The Hohenberg-Kohn theorem and the self-consistent Kohn-Sham scheme 
are straightforwardly extended to the relativistic domain \cite{DG.90}. The 
relativistic Kohn-Sham equation for the auxiliary non-interacting system is 
represented by the single-particle Dirac equation with a local four-potential 
that depends on the ground-state four-current. 
\subsection{Empirical nuclear density functionals}

At the energy and momentum
scales characteristic of nuclei, the only degrees of freedom that have to be
taken into account explicitly in the description of many-body dynamics are
pions and nucleons. The behavior of the nucleon-nucleon (NN) interaction at
long and intermediate distances is determined by one- and two-pion exchange
processes. As already emphasized, short-distance
dynamics cannot be resolved at low energies that
characterize nuclear binding and, therefore, it is represented by
local four-point (contact) NN interactions, with low-energy
(medium-dependent) parameters adjusted to nuclear data. These concepts
of effective field theory and density functional theory methods have recently
been used to derive a microscopic relativistic energy density functional
framework constrained by in-medium QCD sum rules and chiral
symmetry \cite{FKV.04,FKV.06}. The
density dependence of the effective nucleon-nucleon couplings is determined
from the long- and intermediate-range interactions generated by one- and
two-pion exchange processes. They are computed using in-medium chiral
perturbation theory, explicitly including $\Delta(1232)$ degrees of freedom
\cite{FKW.05}. Regularization dependent contributions to the energy density of
nuclear matter, calculated at three-loop level, are absorbed in contact
interactions with parameters representing unresolved short-distance dynamics.

In this work we review a class of relativistic energy density functionals (REDFs)
similar to that introduced in Refs.~\cite{FKV.04,FKV.06} but, instead of using
low-energy QCD constraints for the medium dependence of the parameters,
a phenomenological ansatz is adjusted exclusively to data on nuclear
ground states. This empirical approach,
although guided by microscopic nucleon self-energies in nuclear matter,
gives us more freedom to investigate in detail the relationship between global
properties of a nuclear matter equation of state (volume, surface, and
asymmetry energies) and the corresponding predictions for
properties of finite nuclei.

The basic building blocks of a relativistic nuclear energy density functional are the
densities and currents
bilinear in the Dirac spinor field $\psi$ of the nucleon:
\begin{equation}
\bar{\psi}\mathcal{O}_\tau \Gamma \psi\;, \quad \mathcal{O}_\tau \in \{1,\tau_i\}\;, \quad
   \Gamma \in \{1,\gamma_\mu,\gamma_5,\gamma_5\gamma_\mu,\sigma_{\mu\nu}\}\;.
\end{equation}
Here $\tau_i$ are the isospin Pauli matrices and $\Gamma$ generically denotes
the Dirac matrices. The nuclear ground-state density and energy are determined by the
self-consistent solution of relativistic linear single-nucleon Kohn-Sham equations.
To derive those equations it is useful to construct an
interaction Lagrangian with four-fermion
(contact) interaction terms in the various isospace-space channels:
\begin{center}
\begin{tabular}{ll}
 isoscalar-scalar:   &   $(\bar\psi\psi)^2$\\
 isoscalar-vector:   &   $(\bar\psi\gamma_\mu\psi)(\bar\psi\gamma^\mu\psi)$\\
 isovector-scalar:  &  $(\bar\psi\vec\tau\psi)\cdot(\bar\psi\vec\tau\psi)$\\
 isovector-vector:   &   $(\bar\psi\vec\tau\gamma_\mu\psi)
                         \cdot(\bar\psi\vec\tau\gamma^\mu\psi)$ .\\
\end{tabular}
\end{center}
Vectors in isospin space are denoted by arrows. A general Lagrangian
can be written as a power series in the currents
 $\bar{\psi}\mathcal{O}_\tau\Gamma\psi$ and their derivatives, with higher-order
terms representing in-medium many-body correlations
 \cite{MNH.92,Hoch.94,FML.96,RF.97,BMM.02}.  The problem however, as
 already emphasized, is that the empirical data set of bulk and single-particle
 properties of finite nuclei can only constrain a relatively small set of
 parameters in the general expansion of an effective Lagrangian.
 An alternative, that directly leads to  linear single-nucleon Kohn-Sham equations,
is to construct a Lagrangian with second-order
interaction terms only, with many-body correlations encoded in
density-dependent coupling functions \cite{FKV.04,FKV.06}. In complete analogy to the
successful meson-exchange RMF phenomenology, in which
the isoscalar-scalar $\sigma$ meson, the isoscalar-vector $\omega$ meson,
and the isovector-vector $\rho$ meson build the minimal set of meson fields that is
necessary for a quantitative description of nuclei,
an effective Lagrangian that includes the isoscalar-scalar, isoscalar-vector and
isovector-vector four-fermion interactions reads:
\begin{align}
\label{Lagrangian}
\mathcal{L} &= \bar{\psi} (i\gamma \cdot \partial -m)\psi \nonumber \\
     &- \frac{1}{2}\alpha_S(\hat{\rho})(\bar{\psi}\psi)(\bar{\psi}\psi)
       - \frac{1}{2}\alpha_V(\hat{\rho})(\bar{\psi}\gamma^\mu\psi)(\bar{\psi}\gamma_\mu\psi)
     - \frac{1}{2}\alpha_{TV}(\hat{\rho})(\bar{\psi}\vec{\tau}\gamma^\mu\psi)
                                                                 (\bar{\psi}\vec{\tau}\gamma_\mu\psi) \nonumber \\
    &-\frac{1}{2} \delta_S (\partial_\nu \bar{\psi}\psi)  (\partial^\nu \bar{\psi}\psi)
         -e\bar{\psi}\gamma \cdot A \frac{(1-\tau_3)}{2}\psi\;.
\end{align}
In addition to the free-nucleon Lagrangian and the point-coupling interaction terms,
when applied to nuclei, the model must include the coupling of the protons to the
electromagnetic field. The derivative term in Eq.~(\ref{Lagrangian}) accounts for leading
effects of finite-range interactions that are crucial for a quantitative description of
nuclear density distribution, e.g. nuclear radii. Similar interactions can be included in
each space-isospace channel, but in practice data only constrain a single derivative term,
for instance $\delta_S (\partial_\nu \bar{\psi}\psi)  (\partial^\nu \bar{\psi}\psi) $.
The inclusion of an adjustable derivative term only
in the isoscalar-scalar channel is consistent with conventional meson-exchange
RMF models, in which the mass of the fictitious $\sigma$ meson is adjusted to
nuclear matter and  ground-state properties of finite nuclei, whereas free values are
used for the masses of the $\omega$ and $\rho$ mesons.

The point-coupling Lagrangian Eq.~(\ref{Lagrangian}) does not
include isovector-scalar terms. In the meson-exchange
picture this channel is represented by the exchange of an effective $\delta$ meson,
and its inclusion introduces a proton-neutron effective mass splitting and enhances
the isovector spin-orbit potential.  Although the spin-orbit strength has
a relatively well-defined value, the distribution between the scalar and vector
channels is not determined by ground-state data. To reduce the number of
adjustable parameters, the isovector-scalar channel may be omitted from
an energy density functional that will primarily be used for the description of
low-energy nuclear structure.

In general the strength parameters of the interaction terms in Eq. (\ref{Lagrangian})
are functions of the nucleon 4-current:
\begin{equation}
j^\mu = \bar{\psi} \gamma^\mu \psi = \hat{\rho} u^{\mu} \; ,
\end{equation}
where $u^{\mu}$ is the 4-velocity defined as
$(1-{\bm v}^2)^{-1/2}(1,{\bm v})$. In the rest-frame of homogeneous
nuclear matter: $\bm v=0$. The single-nucleon
Dirac equation, the relativistic analogue of the Kohn-Sham equation, is obtained
from the variation of the Lagrangian with respect to $\bar{\psi}$:
\begin{equation}
\left[ \gamma_\mu(i\partial^\mu - \Sigma^\mu -\Sigma_R^\mu) - (m+\Sigma_S)\right]\psi = 0\;,
\label{Dirac-eq}
\end{equation}
with the nucleon self-energies defined by the following relations:
\begin{align}
\label{sigma_v}
\Sigma^\mu &= \alpha_V(\rho_v) j^\mu + e  \frac{(1-\tau_3)}{2} A^\mu\\
\label{sigma_r}
\Sigma_R^\mu &= \frac{1}{2}\frac{j^\mu}{\rho_v}
            \left\{ \frac{\partial \alpha_S}{\partial \rho}\rho_s^2
         +\frac{\partial \alpha_V}{\partial \rho}j_\mu j^\mu
         + \frac{\partial \alpha_{TV}}{\partial \rho}\vec{j}_\mu \vec{j}^\mu
          \right\}\\
\label{sigma_s}
\Sigma_S &= \alpha_S(\rho_v)\rho_s - \delta_S \Box \rho_s \\
\label{sigma_tv}
\Sigma_{TV}^\mu &=  \alpha_{TV}(\rho_v)\vec{j}^\mu \;.
\end{align}
In addition to the contributions of the isoscalar-vector four-fermion interaction and
the electromagnetic interaction, the isoscalar-vector self-energy $\Sigma^\mu$
includes the ``rearrangement''
terms $\Sigma_R^\mu$, arising from the variation of the vertex
functionals $\alpha_S$, $\alpha_V$, and $\alpha_{TV}$ with respect to the
nucleon fields in the density operator $\hat{\rho}$.
The inclusion of the rearrangement self-energy is essential for
energy-momentum conservation and the thermodynamical consistency of
the model \cite{FLW.95,TW.99,NVFR.02}. $\Sigma_S$  and $\Sigma_{TV}^\mu$
denote the isoscalar-scalar and isovector-vector self-energies, respectively.

In the relativistic density functional framework the nuclear ground state $\rightg$ is
represented by the self-consistent mean-field solution of the system of equations
(\ref{Dirac-eq}) -- (\ref{sigma_tv}), with the
isoscalar and isovector 4-currents and
scalar density:
\begin{eqnarray}
\label{den1}
j_\mu & = \leftg \bar{\psi} \gamma_\mu \psi \rightg =
& \sum_{k=1}^N v_{k}^{2}~\bar{\psi}_k \gamma_\mu \psi_k \; ,\\
\label{den2}
\vec{j}_\mu & =
\leftg \bar{\psi} \gamma_\mu \vec{\tau} \psi \rightg =
& \sum_{k=1}^N v_{k}^{2}~\bar{\psi}_k \gamma_\mu \vec{\tau} \psi_k \; ,\\
\label{den3}
\rho_S & = \leftg \bar{\psi} \psi \rightg =
& \sum_{k=1}^N v_{k}^{2}~\bar{\psi}_k \psi_k \; ,
\end{eqnarray}
where $\psi_k$ are Dirac spinors, and
the sum runs over occupied positive-energy single-nucleon orbitals,
including the corresponding occupation factors $v_{k}^{2}$.
The single-nucleon Dirac equations are solved
self-consistently in the ``no-sea''
approximation that omits the explicit contribution
of negative-energy solutions of the relativistic
equations to the densities and currents. Vacuum polarization effects
are implicitly included in the adjustable density-dependent parameters of the theory.

To determine the density dependence of the coupling functionals
$\alpha_S$, $\alpha_V$, and $\alpha_{TV}$ one
could start from a microscopic (relativistic) equation of state (EoS) of symmetric
and asymmetric nuclear matter, and map the corresponding nucleon self-energies
on the mean-field self-energies Eqs.~(\ref{sigma_v}) - (\ref{sigma_tv}) that
determine the single-nucleon Dirac equation (\ref{Dirac-eq}). This approach has
been adopted, for instance, in RMF models based on Dirac-Brueckner-Hartree-Fock
self-energies in nuclear matter \cite{FLW.95,JL.98,HKL.01}, or on in-medium
chiral perturbation theory (ChPT) calculations of the nuclear matter
EoS \cite{FKV.04,FKV.06}. In general, however, energy density functionals determined
directly from a microscopic EoS do not provide a very accurate description of data
in finite nuclei. The reason, of course, is that a calculation of the nuclear matter EoS
involves approximation schemes and includes adjustable parameters that are not
really constrained by nuclear structure data. The resulting bulk properties of
infinite nuclear matter (saturation density, binding energy, compression modulus,
asymmetry energy) do not determine uniquely the parameters of nuclear
energy density functionals, which usually must be further fine-tuned
to ground-state data (masses and/or charge radii) of spherical nuclei.

In a phenomenological construction of a relativistic energy density functional
one starts from an assumed ansatz for the medium dependence of the mean-field
nucleon self-energies, and adjusts the free parameters directly to ground-states
data of finite nuclei. This procedure was used, for instance, in the construction of
the relativistic density-dependent interactions TW-99~\cite{TW.99}, DD-ME1~\cite{NVFR.02},
DD-ME2~\cite{LNVR.05}, PKDD~\cite{LMGZ.04}, PK01~\cite{Long.06},
DD-PC1 \cite{NVR.08}.

\section{Adjusting parameters to masses: the empirical functional DD-PC1}
\label{DD-PC1}
In an analysis of relativistic nuclear dynamics \cite{PF.06}, modern high-precision
nucleon-nucleon (NN) potentials (Argonne V$_{18}$, Bonn A, CD-Bonn, Idaho, 
Nijmegen, V$_{{\rm low}-k}$) were mapped on a relativistic operator basis, and the 
corresponding relativistic nucleon self-energies in nuclear matter were calculated 
in Hartree-Fock approximation at {\em tree level} \cite{PFD.06}. 
A very interesting result is that, at moderate
nucleon densities relevant for nuclear structure calculations, all potentials
yield very similar scalar and vector mean fields of several hundred MeV
magnitude, in remarkable agreement
with standard RMF phenomenology: at saturation density a large and attractive
scalar field $\Sigma_s \approx -400$ MeV, and a repulsive vector field
$\Sigma_v \approx 350$ MeV. The different treatment of short-distance
dynamics in the various NN potentials leads to slightly more pronounced
differences between the corresponding self-energies at higher nucleon densities.
Generally, however, all potentials predict a very
similar density dependence of the scalar and vector self-energies.
In the chiral effective field theory framework, in particular, these self-energies are
predominantly generated by contact terms that occur at next-to-leading order in
the chiral expansion.

Of course at the Hartree-Fock level these NN potentials do not yield
saturation of nuclear matter. Nevertheless, the corresponding self-energies can
be used as the starting point in the modeling of medium dependence of a
relativistic nuclear energy density functional. The density functional DD-PC1 \cite{NVR.08},
which is representative of the class of semi-empirical REDFs and will be used in
illustrative calculations throughout this work, was adjusted starting from the
Hartree-Fock isoscalar scalar and vector self-energies of the Idaho N$^3$LO
potential \cite{EM.03}. The strength and density dependence of the interaction terms of
the Lagrangian Eq.~(\ref{Lagrangian}) were parameterized by the following ansatz:
\begin{eqnarray}
\alpha_S(\rho)&=& a_S + (b_S + c_S x)e^{-d_S x},\\
\alpha_V(\rho)&=& a_V +  b_V e^{-d_V x},
\label{parameters}
\end{eqnarray}
where $x=\rho/\rho_{sat}$, and $\rho_{sat}$ denotes the nucleon density
at saturation in symmetric nuclear matter.
In the isovector channel the corresponding Hartree-Fock nucleon
self-energies, obtained by directly mapping microscopic NN potentials on a
relativistic operator basis, are presently not available. Therefore, as it was done
in the case of functionals based on finite-range meson-exchange interactions
TW-99 \cite{TW.99}, DD-ME1 \cite{NVFR.02}, DD-ME2 \cite{LNVR.05}, and
PK01 \cite{Long.06}, the density dependence of the isovector-vector coupling
function was modeled based on the results of Dirac-Brueckner calculations of asymmetric
nuclear matter \cite{JL.98}:
\begin{equation}
\alpha_{TV}(\rho)= b_{TV} e^{-d_{TV} x}\; .
\end{equation}

The parameters of a nuclear EDF can be constrained by the choice of the nuclear
matter (symmetric and asymmetric) equation of state. The functional DD-PC1 was
eventually fine-tuned to experimental binding energies of finite nuclei and, since
the calculated nuclear masses are not very sensitive to the nuclear matter saturation
density, this quantity, together with the compression modulus
and the Dirac mass, were kept fixed. The saturation density
$\rho_{sat}=0.152~\textnormal{fm}^{-3}$ is in accordance with values predicted by
most modern relativistic functionals, such as
 DD-ME1 \cite{NVFR.02}, and DD-ME2 \cite {LNVR.05}. From these
functionals also the Dirac effective nucleon mass was taken:
$m^*_D =  m + \Sigma_S = 0.58 m$. The Dirac mass is closely
related to the effective spin-orbit single-nucleon potential, and empirical
energy spacings between spin-orbit partner states in finite nuclei determine
a relatively narrow interval of allowed values: $0.57 \le m^*_D/m \le 0.61$.
In Ref.~\cite{NVLR.08}  it was shown that,
to reproduce experimental excitation energies of isoscalar giant
monopole resonances, point-coupling interactions require a
nuclear matter compression modulus close to 230 MeV,
and thus $K_{\infty} = 230$ MeV was used for DD-PC1.
Nuclear structure data do not constrain the
nuclear matter EoS at high nucleon densities. Therefore, in addition to
$\rho_{sat}$, $m^*_D$, and $K_{\infty}$, two additional points on the
$E(\rho)$ curve in symmetric matter were fixed to the microscopic EoS of Akmal,
Pandharipande and Ravenhall \cite{APR.98}, based on the Argonne V$_{18}$
NN potential and the UIX three-nucleon interaction. This EoS has extensively been
used in studies of high-density nucleon matter and neutron stars.

The isovector channel of the energy density functional determines the
density dependence of the nuclear matter symmetry energy
\begin{equation}
\label{S2}
S_2(\rho) = a_4+\frac{p_0}{\rho_{sat}^2}(\rho-\rho_{sat})
                         + \frac{\Delta K_0}{18 \rho_{sat}^2}(\rho-\rho_{sat})^2+\cdots\;.
\end{equation}
The parameter $p_0$ characterizes the linear density dependence of the
symmetry energy, and $\Delta K_0$ is the isovector correction to
the compression modulus.  Experimental masses, unfortunately, do not
place very strict constraints on the parameters of the expansion
of $S_2(\rho)$ \cite{Fur.02}, but self-consistent mean-field calculations
show that binding energies can restrict the values of
$S_2$ at nucleon densities somewhat
below saturation density, i.e. at $ \rho  \approx 0.1~\textnormal{fm}^{-3}$.
Additional information on the symmetry energy can be obtained from
data on neutron skin thickness and excitation energies of giant dipole
resonances. Although values of neutron radii are available only for a
small number of nuclei and the corresponding uncertainties are
large, recent studies have shown that relativistic effective interactions
with volume asymmetry $a_4$ in the range
$31\;\textnormal{MeV} \le a_4 \le 35\; \textnormal{MeV}$
predict values for neutron skin thickness that are consistent with data,
and reproduce experimental excitation energies of isovector giant dipole resonances
(cf. Ref.~\cite{VNR.03} and references therein). For DD-PC1,
therefore, the volume asymmetry was fixed at $a_4=33\; \textnormal{MeV}$,
and the symmetry energy was varied at a density that corresponds to
an average nucleon density in finite nuclei:
$\langle \rho \rangle =0.12\;\textnormal{fm}^{-3}$.

Until recently the standard procedure of fine-tuning
non-relativistic or relativistic nuclear density functionals, was to perform a
{\em least-squares} adjustment of a given set of free parameters simultaneously
to a favorite nuclear matter EoS and to ground-state properties of about ten to twelve
spherical closed-shell nuclei. Deformed systems have
generally not been included in the adjustment of parameters because calculation of
deformed nuclei is computationally much more intensive, especially in a
multi-parameter fit. Ground-state data of closed-shell nuclei, however, include long-range
correlations that cannot be absorbed into global functionals, e.g. correlations determined 
by the coupling to low-energy collective vibrations. The excitation energies 
and the structure of low-lying collective vibrational modes crucially depend on the details 
of single-nucleon levels in the vicinity of the Fermi surface and, therefore, the 
corresponding ground-state correlations obviously cannot display a smooth 
dependence on nucleon number. On the other hand, it is well known that energy 
density functionals or, at the level of practical application, self-consistent mean-field models  
provide a much better description of deformed, open-shell nuclei. The reason is that the 
mean-field approach includes the mechanism of spontaneous symmetry breaking that 
generates the most important ground-state correlations in deformed nuclei: 
quadrupole correlations in the $ph$-channel and monopole-correlations in 
the $pp$-channel \cite{BHR.03,BK.68}. Additional rotational corrections that arise from the 
restoration of rotational symmetry vary rather smoothly over large mass intervals, and therefore
can be included implicitly in the density functional. The same reasoning applies to  
corrections originating from the restoration of particle number in the regime of strong pairing 
realized in well-deformed nuclei. It also appears that that those features of an effective 
interaction that determine surface properties are better constrained by 
binding energies of deformed systems, as compared to spherical nuclei \cite{Kor.10}. 
The functional DD-PC1 was directly adjusted to binding energies of axially 
symmetric deformed nuclei in the mass regions $A\approx 150-180$ and $A\approx 230-250$. 
Similar approaches, employing data on both spherical and deformed nuclei, have recently been 
adopted in the optimization of nuclear energy density functionals of Skyrme type 
\cite{Kor.10,GSP.09,KRBM.09}. 

Calculated masses of finite
nuclei are primarily sensitive to the three leading terms in the empirical mass formula:
volume, surface and symmetry energy
\begin{equation}
\label{SEMF}
B.E. = a_v A +a_s A^{2/3}+a_4\frac{(N-Z)^2}{4A}+\cdots\;,
\end{equation}
where $a_v$, $a_s$ and $a_4$ correspond to the volume binding energy, surface
energy, and symmetry energy, respectively, at saturation density in nuclear matter.
One can, therefore, generate families of effective interactions that are characterized
by different values of $a_v$, $a_s$ and $a_4$ (or symmetry energy at a lower density,
as explained above), and determine which parametrization
minimizes the deviation from the empirical binding energies. Of course, if a functional
is adjusted by varying the volume, symmetry, and surface energies, the parameters that
determine these quantities will generally be correlated because of
Eq.~(\ref {SEMF}). When only a small number of nuclei is considered, satisfactory
results can be obtained with various, in general linearly dependent
combinations of parameters. The parameters of the functional DD-PC1 were thus
determined in a careful comparison of the predicted binding energies with data, for a
set of 64 nuclei with $A\approx 150-180$ and $A\approx 230-250$.

\begin{figure}
\begin{center}
\includegraphics[scale=0.625]{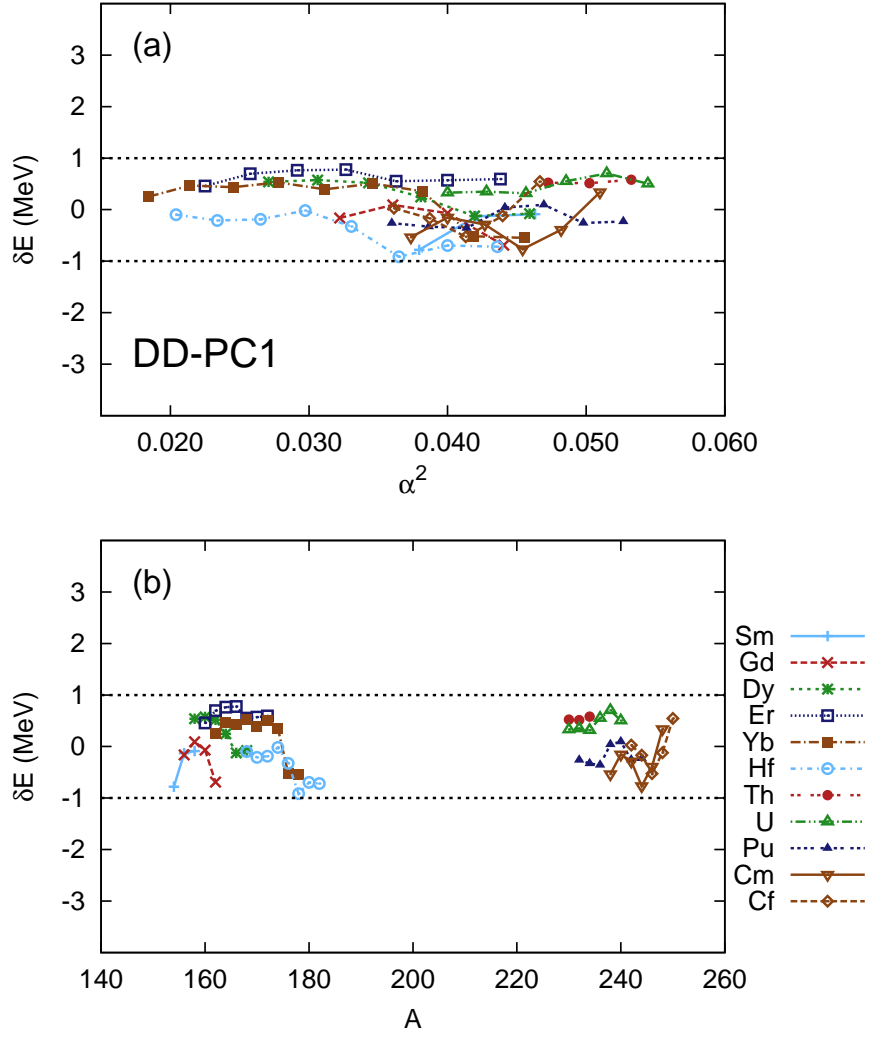}
\end{center}
\caption{
Absolute deviations of the calculated binding energies from the
experimental values of the 64 axially deformed nuclei,
as functions of the asymmetry coefficient (upper panel), and
mass number (lower panel). Lines connect nuclei that
belong to the isotopic chains shown in the legend.
The theoretical binding energies are calculated using the
relativistic density functional DD-PC1.}
\label{Fig_residuals}
\end{figure}

The ground states of 64 axially deformed nuclei were calculated in the self-consistent
mean field approximation. Pairing correlations were treated in the BCS constant-gap
approximation with empirical pairing gaps (5-point formula), and the pairing model 
space included two major oscillator shells ($2\hbar\omega_0$) above the Fermi surface.
This approximation is justified because pairing correlations contribute only a very small 
portion to the total binding energy. In nuclei there is a clear separation of scales between the
bulk contributions to the binding energies of the order of hundreds to more than thousand MeV, 
and the pairing energy of the order of ten MeV. To take into account pairing correlations 
in a calculation of the binding of nuclei close to $\beta$-stability, therefore, 
 it is sufficient to consider only the monopole part of the effective pairing interaction adjusted 
to experimental paring gaps. This is, of course, no longer true in studies of phenomena 
determined by structure effects in the vicinity of the Fermi surface, such as nuclear excitations 
or fission barriers, or in nuclei far from stability, where detailed properties of the effective 
interaction in the pairing channel become important. In this review we will also present 
examples with more realistic pairing interactions, e.g. a zero-range force or a separable 
version of the Gogny force.

A careful analysis of deviations between calculated and experimental masses
(mass residuals),  showed a pronounced
isospin and mass dependence of the residuals on the nuclear matter volume
energy at saturation. To reduce the absolute mass residuals to less than
1 MeV, and to contain their mass and isotopic dependence, strict constraints
on the value of $a_v$ must be met. The narrow window of allowed values
of the volume energy cannot be determined microscopically already at the
nuclear matter level, but rather results from a fine-tuning of the parameters
of the energy density functional to experimental masses.
Calculated binding energies and charge radii are also sensitive to the
choice of the surface coefficient $a_s$ that determines the
surface energy and surface thickness of semi-infinite nuclear matter.
For the optimal density functional DD-PC1, characterized by the
following properties at the saturation point: nucleon density
$\rho_{sat}=0.152~\textnormal{fm}^{-3}$, volume energy $a_v=-16.06$
MeV, surface energy $a_s=17.498$ MeV, symmetry energy $a_4 = 33$ MeV,
and the nuclear matter compression modulus $K_{nm} = 230$ MeV,
in Fig.~\ref{Fig_residuals} we display the absolute deviations of the
calculated binding energies from the experimental values as functions of the
isospin asymmetry
\begin{displaymath}
\label{alpha}
\alpha^2 = \frac{(N-Z)^2}{A^2}\; ,
\end{displaymath}
and nucleon number. Positive deviations correspond to under-bound nuclei.
The functional DD-PC1 corresponds to the lowest $\chi^2$ value in the multi-parameter
fit, and does not display any visible isotopic or mass dependence of the
deviations of calculated masses. The absolute errors for
all 64 axially deformed nuclei in the mass regions $A\approx 150-180$
and $A\approx 230-250$ are smaller than $1$~MeV. With stronger
binding in symmetric nuclear matter (i.e. by increasing the
absolute value of $a_v$), the corresponding deviations of calculated
binding energies become larger, and they also
acquire a definite isotopic dependence. Reducing the absolute value of
$a_v$ reverses the isotopic trend of the errors.

The density functional DD-PC1 has been further tested
in calculations of properties of spherical and deformed medium-heavy and heavy
nuclei, including binding energies, charge radii, deformation parameters, neutron skin
thickness, and excitation energies of giant multipole resonances. Results have been
compared with available data, and with predictions of the most successful
finite-range meson-exchange relativistic effective interactions. In general,
a very good agreement with data has been obtained except, perhaps, for the effect
of overbinding of spherical closed-shell nuclei. DD-PC1, like virtually all relativistic
mean-field models, is characterized by a relatively low effective nucleon mass and,
when adjusted to masses of deformed nuclei, it overbinds spherical closed-shell
systems. The well known problem of ``arches'' of mass residuals between shell
closures could, in principle, be addressed by a functional that goes beyond the static
mean-field approximation and includes an explicit energy dependence of the nucleon
self-energies. Very good results have been obtained for the excitation energies
of giant monopole and dipole resonances in spherical nuclei, calculated using
the relativistic quasiparticle random-phase approximation based on the DD-PC1
functional. The agreement with data validates the choice of the nuclear matter
compressibility and symmetry energy for DD-PC1.
The total number of parameters is 10, similar to most non-relativistic Skyrme-type
density functionals. The effective Lagrangian of
DD-PC1 contains only four interaction terms except, of course, the Coulomb term
(cf. Eq.~(\ref{Lagrangian})), and the 10 parameters determine the density dependence
of the strength functionals and reflect the complex nuclear many-body dynamics.

In the following sections we will also employ DD-PC1 in a series of spectroscopic 
calculations utilizing models that go beyond the ``mean-field" level, and explicitly 
include collective correlations related to restoration of broken symmetries and 
fluctuations in quadrupole coordinates. Strictly speaking, those correlations 
that are treated explicitly should not be included in the density functional 
in an implicit way, that is, in the parameters adjusted to data that already 
include correlations. In future studies the solution 
could be to adjust the functional to pseudodata, obtained by subtracting 
correlation effects from experimental masses and, eventually, radii. 
In deformed nuclei the dominant contribution to ground-state correlations 
is the rotational energy correction \cite{BBeH.06}, which is relatively simple to calculate. 
Approximate methods have been developed that enable a systematic evaluation 
of correlation energies for the nuclear mass table \cite{BBH.04}. Starting from a 
set of pseudodata, one expects that the corresponding modifications of the parameters 
of the energy density functional will be relatively small. However, even a small 
change in the relative contribution of various interaction terms could be  
the decisive factor in specific cases of soft potential energy surfaces, 
coexistence of prolate and oblate shapes, level ordering, etc. 
\section{3D Relativistic Hartree-Bogoliubov model with a separable pairing interaction}
\label{RHB}
Relativistic energy density functionals have been employed in
analyses of properties of ground and excited states in spherical and deformed
nuclei. For a quantitative analysis of open-shell nuclei it is necessary to
consider also pairing correlations. Pairing has often been taken into account
in a very phenomenological way in the BCS model with the monopole pairing
force, adjusted to the experimental odd-even mass differences. In many cases,
however, this approach presents only a poor approximation. The physics of
weakly-bound nuclei, in particular, necessitates a unified and self-consistent
treatment of mean-field and pairing correlations. This has led to the
formulation and development of the relativistic Hartree-Bogoliubov (RHB)
model~\cite{Ring.96}, which represents a relativistic extension of the
conventional Hartree-Fock-Bogoliubov framework. The RHB model provides a
unified description of particle-hole $(ph)$ and particle-particle $(pp)$
correlations on a mean-field level by using two average potentials: the
self-consistent mean field that encloses all the long range \textit{ph}
correlations, and a pairing field $\hat{\Delta}$ which sums up the
\textit{pp}-correlations. The ground state of a nucleus is described by a
generalized Slater determinant $|\Phi\rangle$ that represents the vacuum with
respect to independent quasiparticles. The quasiparticle operators are defined
by the unitary Bogoliubov transformation of the single-nucleon creation and
annihilation operators:
\begin{equation}
\alpha_{k}^{+}=\sum\limits_{l}U_{lk}^{{}}c_{l}^{+}+V_{lk}^{{}}c_{l}^{{}}\;,
\end{equation}
where $U$ and $V$ are the Hartree-Bogoliubov wave functions determined by the
solution of the RHB equation. In coordinate representation:
\begin{equation}
\label{eq:RHB}\left(
\begin{array}
[c]{cc}%
h_{D}-m-\lambda & \Delta\\
-\Delta^{*} & -h_{D}^{*}+m+\lambda
\end{array}
\right)  \left(
\begin{array}
[c]{c}%
U_{k}({\mbox{\boldmath $r$}})\\
V_{k}({\mbox{\boldmath $r$}})
\end{array}
\right)  = E_{k} \left(
\begin{array}
[c]{c}%
U_{k}({\mbox{\boldmath $r$}})\\
V_{k}({\mbox{\boldmath $r$}})
\end{array}
\right)  \;.
\end{equation}
In the relativistic case the self-consistent mean-field corresponds to the
single-nucleon Dirac Hamiltonian $\hat{h}_{D}$ of Eq.~(\ref{Dirac-eq}).
$m$ is the nucleon mass, and the chemical potential
$\lambda$ is determined by the particle number subsidiary condition such
that the expectation value of the particle number operator in the ground state
equals the number of nucleons. The pairing field $\Delta$ reads
\begin{equation}
\Delta_{ab}({\mbox{\boldmath $r$}},{\mbox{\boldmath $r$}}^{\prime}) = \frac
{1}{2}\sum_{c,d}{V_{abcd}({\mbox{\boldmath $r$}},{\mbox{\boldmath $r$}}%
^{\prime})} \kappa_{cd}({\mbox{\boldmath $r$}},{\mbox{\boldmath $r$}}^{\prime
}).
\end{equation}
where $V_{abcd}({\mbox{\boldmath $r$}},{\mbox{\boldmath $r$}}^{\prime})$ are
the matrix elements of the two-body pairing interaction, and the indices $a$,
$b$, $c$ and $d$ denote the quantum numbers that specify the Dirac indices of
the spinor. The column vectors denote the quasiparticle wave functions, and
$E_{k}$ are the quasiparticle energies. The dimension of the RHB matrix
equation is two times the dimension of the corresponding Dirac equation. For
each eigenvector $(U_{k} ,V_{k} )$ with positive quasiparticle energy $E_{k} >
0$, there exists an eigenvector $(V_{k}^{*},U_{k}^{*})$ with quasiparticle
energy $-E_{k}$. Since the baryon quasiparticle operators satisfy fermion
commutation relations, the levels $E_{k}$ and $-E_{k}$ cannot be occupied
simultaneously. For the solution that corresponds to a ground state of a
nucleus with even particle number, one usually chooses the eigenvectors with
positive eigenvalues $E_{k}$.

The single-particle density and the pairing tensor, constructed from the
quasi-particle wave functions
\begin{align}
\label{eq:pairing-tensor}\rho_{cd}({\mbox{\boldmath $r$}}%
,{\mbox{\boldmath $r$}}^{\prime})  &  = \sum_{k>0}{V^{*}_{ck}%
({\mbox{\boldmath $r$}})V_{dk}({\mbox{\boldmath $r$}}^{\prime})},\\
\kappa_{cd}({\mbox{\boldmath $r$}},{\mbox{\boldmath $r$}}^{\prime})  &
=\sum_{k>0}{U^{*}_{ck}({\mbox{\boldmath $r$}})V_{dk}({\mbox{\boldmath $r$}}%
^{\prime})},
\end{align}
are calculated in the \emph{no-sea} approximation (denoted by $k>0$): the
summation runs over all quasiparticle states $k$ with positive quasiparticle
energies $E_{k}>0$, but omits states that originate from the Dirac sea. The
latter are characterized by quasiparticle energies larger than the Dirac gap
($\approx1200$ MeV).

Pairing correlations in nuclei are restricted to an energy window of a few MeV
around the Fermi level, and their scale is well separated from the scale of
binding energies, which are in the range of several hundred to thousand MeV.
There is no empirical evidence for any relativistic effect in the nuclear
pairing field $\hat{\Delta}$ and, therefore, a hybrid RHB model with a
non-relativistic pairing interaction can be employed. For a general two-body
interaction, the matrix elements of the relativistic pairing field read
\begin{equation}
\hat{\Delta}_{a_{1} p_{1}, a_{2} p_{2}} = {\frac{1}{2}}\sum\limits_{a_{3}
p_{3}, a_{4} p_{4}} \langle a_{1} p_{1}, a_{2} p_{2} |V^{pp}|a_{3} p_{3},
a_{4} p_{4}\rangle_{a}~ \kappa_{a_{3} p_{3}, a_{4} p_{4}}\; ,
\end{equation}
where the indices ($p_{1},p_{2},p_{3},p_{4} \equiv f, g$) refer to the large
and small components of the quasiparticle Dirac spinors:
\begin{equation}
U(\mathbf{r},s,t)\ =\ \left(
\begin{array}
[c]{c}%
f_{U}(\mathbf{r},s,t)\\
ig_{U}(\mathbf{r},s,t)
\end{array}
\right)  \quad\quad\quad V(\mathbf{r},s,t)\ =\ \left(
\begin{array}
[c]{c}%
f_{V}(\mathbf{r},s,t)\\
ig_{V}(\mathbf{r},s,t)
\end{array}
\right)  \; . \label{UV}%
\end{equation}
In practical applications of the RHB model to finite open-shell nuclei, only
the large components of the spinors $U_{k}(\mathbf{r})$ and $V_{k}%
(\mathbf{r})$ are used to build the non-relativistic pairing tensor
$\hat{\kappa}$ in Eq.~(\ref{eq:pairing-tensor}). The resulting pairing field
reads
\begin{equation}
\hat{\Delta}_{a_{1} f, a_{2} f} = {\frac{1}{2}}\sum\limits_{a_{3} f, a_{4} f}
\langle a_{1} f, a_{2} f |V^{pp}|a_{3} f, a_{4} f\rangle_{a}~ \kappa_{a_{3} f,
a_{4} f}\; .
\end{equation}
The other components: $\hat{\Delta}_{fg}$, $\hat{\Delta}_{gf}$, and
$\hat{\Delta}_{gg}$ can be safely omitted~\cite{SR.02}.

In most applications of the RHB model~\cite{VALR.05} the pairing part of the
Gogny force~\cite{BGG.84} has
been employed in the particle-particle ($pp$) channel:
\begin{equation}
V^{pp}(1,2)~=~\sum_{i=1,2}e^{-((\mathbf{r}_{1}-\mathbf{r}_{2})/{\mu_{i}})^{2}%
}\,(W_{i}~+~B_{i}P^{\sigma}-H_{i}P^{\tau}-M_{i}P^{\sigma}P^{\tau})\;,
\label{Gogny}%
\end{equation}
with the set D1S~\cite{BGG.91} for the parameters $\mu_{i}$, $W_{i}$, $B_{i}$,
$H_{i}$, and $M_{i}$ $(i=1,2)$. A basic advantage of the Gogny force is the
finite range, which automatically guarantees a proper cut-off in momentum
space. However, the resulting pairing field is non-local and the solution of
the corresponding Dirac-Hartree-Bogoliubov integro-differential equations can
be time-consuming, especially in the case 3D calculations for nuclei with
triaxial shapes.

In a series of recent articles~\cite{TMR.09a,TMR.09b,TMR.09c,NRV.10}
a separable form of the pairing force has been introduced for RHB calculations
in spherical and deformed nuclei. The force is separable in momentum
space, and is completely determined by two parameters that are adjusted to
reproduce in symmetric nuclear matter the bell-shape curve of the pairing gap
of the Gogny force. The gap equation in the $^{1}$S$_{0}$
channel reads
\begin{equation}
\Delta(k)=-\int_{0}^{\infty}{\frac{{k^{\prime2}dk^{\prime}}}{{2\pi^{2}}}%
}\left\langle k\right\vert V^{^{1}S_{0}}\left\vert k^{\prime}\right\rangle
{\frac{{\Delta(k^{\prime})}}{{2E(k^{\prime})}}}\;,
\end{equation}
and the pairing force is separable in momentum space:
\begin{equation}
\left\langle k\right\vert V^{^{1}S_{0}}\left\vert k^{\prime}\right\rangle
=-Gp(k)p(k^{\prime})\;. \label{sep_pair}%
\end{equation}
By assuming a simple Gaussian ansatz $p(k)=e^{-a^{2}k^{2}}$, the two
parameters $G$ and $a$ have been adjusted to reproduce the density dependence
of the gap at the Fermi surface, calculated with a Gogny force. For the D1S
parameterization~\cite{BGG.91} of the Gogny force the following
values were determined: $G=-728\;\mathrm{MeVfm}%
^{3}$ and $a=0.644\;\mathrm{fm}$. When the pairing force Eq.~(\ref{sep_pair})
is transformed from momentum to coordinate space, it takes the form:
\begin{equation}
V({\mbox{\boldmath $r$}}_{1},{\mbox{\boldmath $r$}}_{2},{\mbox{\boldmath $r$}}%
_{1}^{\prime},{\mbox{\boldmath $r$}}_{2}^{\prime})=G\delta\left(
{\mbox{\boldmath $R$}}-{\mbox{\boldmath $R$}}^{\prime}\right)
P({\mbox{\boldmath $r$}})P({\mbox{\boldmath $r$}}^{\prime})\frac{1}{2}\left(
1-P^{\sigma}\right)  , \label{pp-force}%
\end{equation}
where ${\mbox{\boldmath $R$}}=\frac{1}{2}\left(  {\mbox{\boldmath $r$}}%
_{1}+{\mbox{\boldmath $r$}}_{2}\right)  $ and ${\mbox{\boldmath $r$}}%
={\mbox{\boldmath $r$}}_{1}-{\mbox{\boldmath $r$}}_{2}$ denote the
center-of-mass and the relative coordinates, and $P({\mbox{\boldmath $r$}})$
is the Fourier transform of $p(k)$:
\begin{equation}
P({\mbox{\boldmath $r$}})=\frac{1}{\left(  4\pi a^{2}\right)  ^{3/2}%
}e^{-{\mbox{\boldmath $r$}}^{2}/4a^{2}}\;. \label{P3D}%
\end{equation}
The pairing force has finite range and, because of the presence of the factor
$\delta\left(  {\mbox{\boldmath $R$}}-{\mbox{\boldmath
$R$}}^{\prime}\right)  $, it preserves translational invariance. Even though
$\delta\left(  {\mbox{\boldmath $R$}}-{\mbox{\boldmath
$R$}}^{\prime}\right)  $ implies that this force is not completely separable
in coordinate space, the corresponding anti-symmetrized $pp$ matrix elements
\begin{equation}
\left\langle \alpha\bar{\beta}\right\vert V\left\vert \gamma\bar{\delta}
\right\rangle _{a}=\left\langle \alpha\bar{\beta}\right\vert V\left\vert
\gamma\bar{\delta}\right\rangle -\left\langle \alpha\bar{\beta}\right\vert
V\left\vert \bar{\delta}\gamma\right\rangle \;,
\label{eq:matris-element-antisymmetric}%
\end{equation}
can be represented as a sum of a finite number of separable terms in the basis
of a 3D harmonic oscillator.

The Dirac-Hartree-Bogoliubov equations (\ref{eq:RHB}) are solved by expanding the nucleon
spinors in the basis of a 3D harmonic oscillator in Cartesian coordinates \cite{NRV.10}.
In this way both axial and triaxial nuclear shapes can be described. To obtain complete convergence,
for medium heavy nuclei the basis must include at least $N_f^{max} = 14$ major oscillator shells, and
for very heavy nuclei $N_f^{max} = 16$ major oscillator shells are necessary.
The map of the energy surface as a function of the quadrupole deformation is obtained
by imposing constraints on the axial and triaxial quadrupole moments. The method of
quadratic constraints uses an unrestricted variation of the function
\begin{equation}
\label{eq:quadrupole-constraints1}
\langle \hat{H}\rangle +\sum_{\mu=0,2}{C_{2\mu}
   \left(\langle \hat{Q}_{2\mu}\rangle -q_{2\mu} \right)^2},
\end{equation}
where $\langle \hat{H}\rangle$ is the total energy, and $\langle \hat{Q}_{2\mu}\rangle$
denotes the expectation value of the mass quadrupole operators:
\begin{equation}
\label{eq:quadrupole-constraints2}
\hat{Q}_{20}=2z^2-x^2-y^2 \quad \textrm{and} \quad \hat{Q}_{22}=x^2-y^2 \;.
\end{equation}
$q_{2\mu}$ is the constrained value of the multipole moment, and $C_{2\mu}$
the corresponding stiffness constant \cite{RS.80}. 
For a self-consistent solution the quadratic
constraint adds an extra force $\sum_{\mu=0,2}\lambda_{\mu}\hat{Q}_{2\mu}$ to
the system, where $\lambda_{\mu}=2C_{2\mu}(\langle\hat{Q}_{2\mu}\rangle
-q_{2\mu})$. Such a force is necessary to keep the system at a point
different from a stationary point. In general, the values of the quadrupole 
moments $\langle\hat{Q}_{2\mu}\rangle$ of the self-consistent solution coincide 
with the constrained values $q_{2\mu}$  only at the stationary points. Moreover, 
the difference between the quadrupole moment $\langle\hat{Q}_{2\mu}\rangle$
and the constrained value $q_{2\mu}$ depends on the value of the stiffness constant, that 
is, smaller values of $C_{2\mu}$ lead to larger deviations of the quadrupole moment
from the corresponding constrained value. Increasing the value of the stiffness constant, 
however, often destroys the convergence of the self-consistent procedure. This deficiency 
can be resolved by implementing the augmented Lagrangian method \cite{SSBN.10}, 
and this approach has been used in all constrained calculations presented in 
this work.

\begin{figure}
\begin{center}
\includegraphics[scale=0.45]{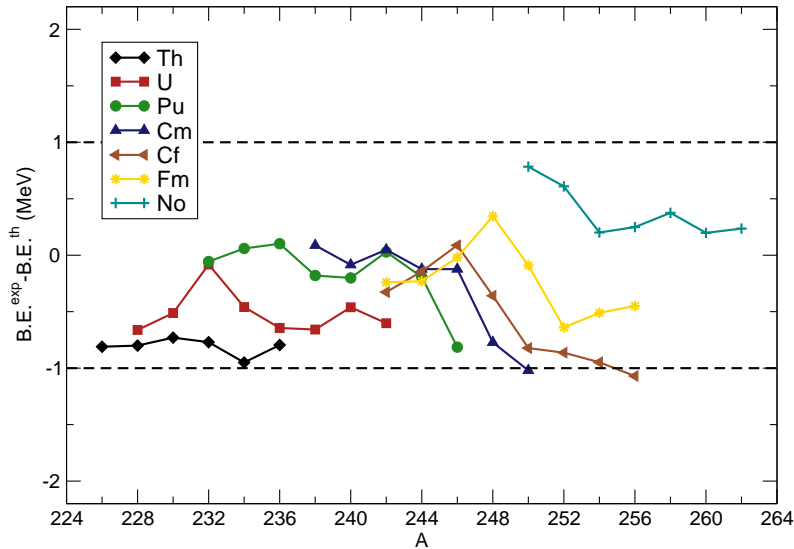}
\end{center}
\caption{Absolute deviations of the calculated relativistic Hartree-Bogoliubov 
(RHB) binding energies from the experimental
values \protect \cite{AW.03}, for the isotopic chains of Th, U, Pu,
Cm, Cf, Fm, and No.}
\label{actinides_m}
\end{figure}
\begin{figure}[t]
\begin{center}
\includegraphics[scale=0.525]{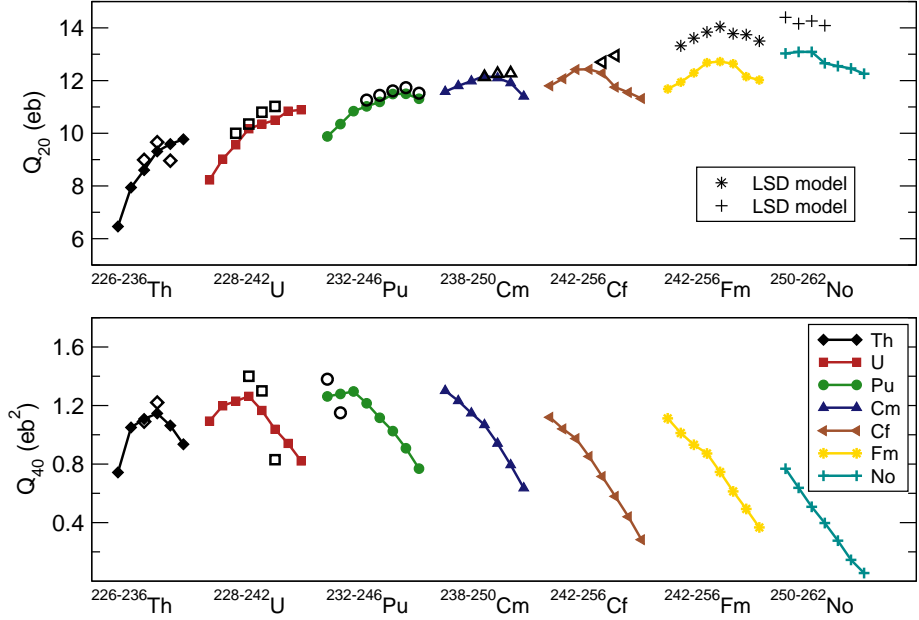}
\end{center}
\caption{Self-consistent RHB ground-state axial quadrupole  and hexadecapole
moments in comparison with data \protect \cite{RNT.01} (open symbols),  for
the isotopic chains of Th, U, Pu, Cm, Cf, Fm, and No. For Fm and No the
calculated quadrupole moments are compared with values predicted by the LSD
model  \protect \cite{LSD.05}.}
\label{actinides_q}
\end{figure}

Figs.~\ref {actinides_m} and \ref{actinides_q}  display the results of self-consistent 3D RHB
calculations for the isotopic chains of Th, U, Pu, Cm, Cf, Fm, and No. The absolute
deviations of the calculated binding energies from data \cite{AW.03}
(Fig.~\ref {actinides_m}) show an excellent agreement between theory and experiment.
Except for the isotopes of No, all the calculated nuclei are slightly underbound,
but the absolute differences between calculated and experimental binding energies
is less than 1 MeV in all cases. The calculated ground-state quadrupole $Q_{20}$, and
hexadecapole $Q_{40}$, moments are compared with available data \cite{RNT.01}
in Fig.~\ref{actinides_q}. We notice that the values predicted by the DD-PC1 functional reproduce
in detail the isotopic trend of the empirical moments in the Th, U, Pu and Cm sequences,
and are in very good agreement with the quadrupole moments of the Cf isotopes.
No data on ground-state moments are available for the isotopes of Fm and No. The
calculated 3D RHB quadrupole moments are compared
with values estimated in axially-symmetric calculations with the Lublin-Strasburg drop
(LSD) model \cite{LSD.05}.

\begin{figure}[htb]
\begin{center}
\includegraphics[scale=0.6]{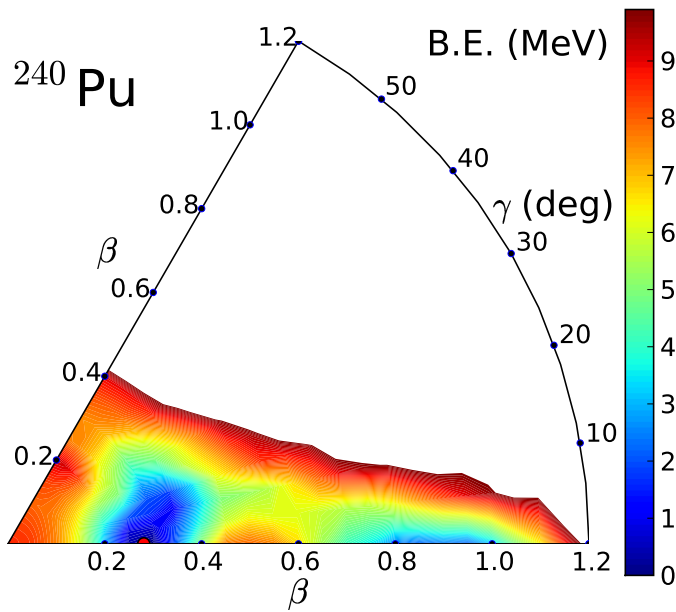}
\end{center}
\caption{\label{PES-240Pu} Self-consistent RHB triaxial
quadrupole binding energy maps of  $^{240}$Pu in the $\beta-\gamma$ plane
($0\le \gamma \le 60^0$). All energies are normalized with respect to the binding
energy of the absolute minimum. The color code refers to the energy
of each point on the surface relative to the minimum.}
\end{figure}

\begin{figure}[htb]
\begin{center}
\includegraphics[scale=0.5]{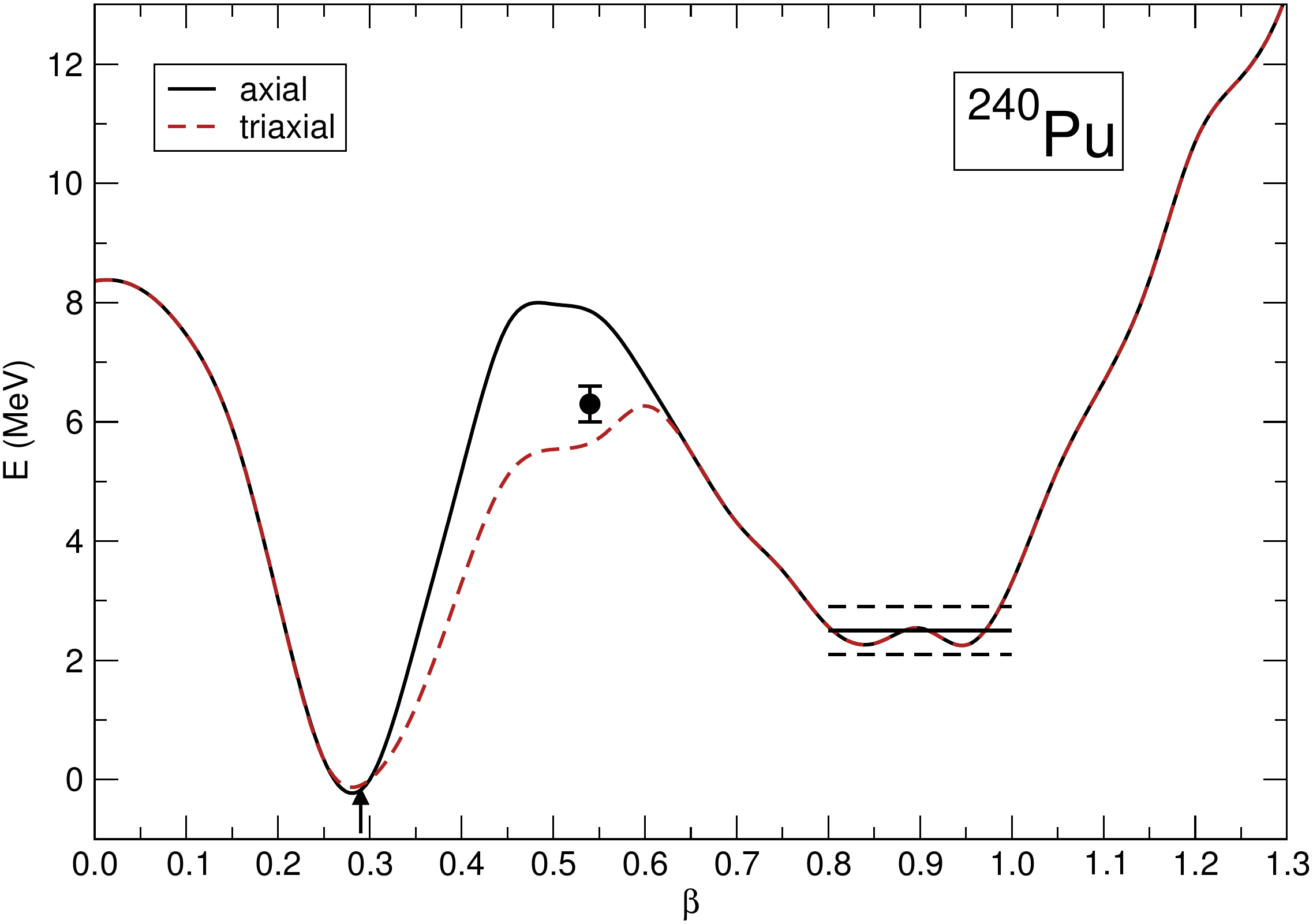}
\end{center}
\caption{\label{PEC-240Pu}(Color online) Deformation energy curves and
the inner barrier of $^{240}$Pu as functions of the axial deformation $\beta$.
The two curves correspond to the axially-symmetric RHB calculation (solid),
and to the projection on the $\beta$-axis of the triaxial PES (dashed),
calculated with the functional DD-PC1.
The experimental values for the ground-state deformation, the barrier
height, and the energy of the second minimum are indicated, respectively, with an
arrow, a symbol with error bars and three lines indicating the value and its errors.
The data are taken from Refs. \protect\cite{BJO80,FIR96,LOB70,BEM73}.}
\end{figure}

The structure of the nucleus $^{240}$Pu and its double-humped fission barrier
has become a standard benchmark for models based on the self-consistent
mean-field approach and the corresponding effective interactions or density
functionals. In Fig.~\ref{PES-240Pu} we display the RHB triaxial quadrupole binding
energy map of $^{240}$Pu in the $\beta - \gamma$ plane ($0\le \gamma\le 60^0$),
calculated with the DD-PC1 energy density functional plus the pairing interaction
Eq.~(\ref{pp-force}) \cite{Li.10b}. The calculation has been carried out on a mesh of
quadrupole deformation parameters with  $\Delta \beta = 0.05$
and $\Delta \gamma = 6^0$. All energies are normalized with respect to
the binding energy of the absolute minimum, and the color code refers to the energy
of each point on the surface relative to the minimum. Since the present implementation
of the model does not include reflection asymmetric shapes, the potential energy
surface (PES) is calculated only up to $\beta \leq 1.3$. For larger deformations,
i.e. in the region of the second barrier, octupole deformations should be taken
into account. The absolute minimum is calculated
at  $\beta = 0.28$, $\gamma = 0^0$, and a second (super-deformed) valley is
predicted around  $\beta \approx 0.9$. The axially symmetric barrier at
$\beta \approx 0.5$ is bypassed through the triaxial region, bringing the height
of the barrier much closer to the empirical value. This is shown more clearly in
Fig.~\ref{PEC-240Pu}, where we plot the deformation energy curves and
the inner barrier of $^{240}$Pu as functions of the axial deformation $\beta$.
The two curves correspond to the axially-symmetric RHB calculation (solid),
and to the projection on the $\beta$-axis of the triaxial PES (dashed).
The experimental values for the ground-state deformation, the barrier
height, and the energy of the second minimum are taken from
Refs.~\cite{BJO80,FIR96,LOB70,BEM73}. One might notice a
very good agreement between theory and available data. In particular, the inclusion
of triaxial shapes lowers the inner barrier by $\approx$ 2 MeV. Similar results
have also been obtained in constrained self-consistent mean-field calculations
using Skyrme functionals \cite{BHR.03}, and in the HFB+Gogny analysis of the
actinide region \cite{DGGL.06} it was shown that the inner barriers of the actinides
are systematically lowered by up to 4 MeV when calculations included triaxial shapes.

The 3D relativistic Hartree-Bogoliubov model, with the functional DD-PC1 in the
particle-hole channel and a separable pairing force in the particle-particle
channel, enables very efficient constrained self-consistent triaxial calculations
of binding energy maps as functions of quadrupole deformation
in the $\beta - \gamma$ plane. The resulting single-quasiparticle energies
and wave functions can be employed as microscopic input for
the generator coordinate method configuration mixing of angular-momentum
projected triaxial wave functions, or can be used to determine
the parameters of the collective Hamiltonian for vibrations and rotations:
the mass parameters, the moments of inertia, and the collective potential.
The solution of the corresponding eigenvalue problem yields the excitation
spectra and collective wave functions that are used in the calculation of
electromagnetic transition probabilities. This approach will be illustrated in
the next two sections.

\section{Beyond the mean-field approximation: restoring broken symmetries and
configuration mixing calculations}
\label{GCM}
Nuclear structure far from stability has become a subject
of extensive experimental and theoretical studies.
The variation of ground-state shapes in an isotopic chain, for
instance, is governed by the evolution of shell structure.
Far from the $\beta$-stability line, in particular, the
energy spacings between single-particle levels change considerably
with the number of neutrons and/or protons. This can result in
reduced spherical shell gaps, modifications of shell structure,
and in some cases spherical magic numbers may disappear.
The reduction of a spherical shell closure is associated with the
occurrence of deformed ground states and, in a number of cases,
with the phenomenon of shape coexistence.

A quantitative description of structure phenomena related to shell evolution
necessitates the inclusion of many-body correlations beyond the mean-field
approximation. The starting point is usually a constrained Hartree-Fock
plus BCS (HFBCS), or Hartree-Fock-Bogoliubov (HFB) calculation of the
potential energy surface with the mass quadrupole components as
constrained quantities. When based on microscopic EDFs or effective
interactions, such calculations comprise short- and long-range many-body
correlations, and result in static symmetry-breaking product many-body states.
Static mean-field models provide a description of bulk properties,
such as masses and radii.
To calculate energy spectra and transition probabilities, however, it is essential
to include collective correlations that arise from symmetry restoration and
fluctuations around the mean-field minimum. Because they are susceptible
to shell effects and vary with nucleon number, collective correlations cannot
be be incorporated in a universal EDF. In deformed nuclei, for instance,
the rotational energy correction, that is, the energy gained by the restoration
of rotational symmetry, is proportional to the quadrupole deformation of
the symmetry-breaking state and can reach several MeV for a well deformed
configuration. Fluctuations of quadrupole deformation also contribute to the
correlation energy. Both types of correlations can be included
simultaneously by mixing angular-momentum projected states corresponding
to different quadrupole moments. The most effective approach for
configuration mixing calculations is the generator coordinate method (GCM),
with multipole moments used as coordinates that generate the intrinsic wave
functions.

In recent years several accurate and efficient models and algorithms,
based on microscopic density functionals or effective interactions,
have been developed that perform the restoration of symmetries broken
by the static nuclear mean field, and take into account quadrupole fluctuations.
Many interesting phenomena related to shell evolution have been investigated by
employing the angular-momentum projected GCM with the axial quadrupole moment
as the generating coordinate, and with intrinsic configurations calculated
in the HFB model with the finite-range Gogny interaction
\cite{RER.02a,GER.02,RER.04,RE.07,RE.08}. Very recently this approach has been
extended to include full triaxial angular-momentum and particle-number
projection \cite{RE.10}. Another sophisticated structure model that takes
into account collective correlations is based on axially constrained HF+BCS
calculations with Skyrme effective interactions in the particle-hole
channel and a density-dependent contact force in the pairing channel
\cite{VHB.00,BFH.03,DBBH.03,BBDH.04,BBH.06}.
Particle numbers and rotational symmetry are restored by projecting
self-consistent mean-field wave functions on the correct numbers
of neutrons and protons, and on angular momentum. Finally, a mixing
of the projected wave functions corresponding to different quadrupole
moments is performed with a discretized version of the generator
coordinate method. The latest extension of this model that incorporates
triaxial angular-momentum projection was reported in Ref.~\cite{BH.08}.

In a series of recent articles we have expanded the framework of
relativistic energy density
functionals to include correlations related to the restoration of
broken symmetries and to fluctuations of collective variables. A
model has been developed that uses the GCM to perform configuration
mixing of angular-momentum \cite{NVR.06a}, and also
particle-number projected \cite{NVR.06b} relativistic
wave functions. The geometry was restricted to axially symmetric
shapes, and the intrinsic wave functions were generated from
solutions of the relativistic mean-field + Lipkin-Nogami BCS
equations, with a constraint on the mass quadrupole moment.
In the first application~\cite{PRL99}, the GCM based on relativistic
EDFs was employed in a study of shape transitions in Nd
isotopes. This approach has been further developed in
Refs.~\cite{Yao.09,Yao.10} by implementing a model that includes
triaxial angular-momentum projection.

Here we will illustrate the ``beyond mean-field" extension of the relativistic
EDF approach by considering a rather simple example of a GCM configuration
mixing of angular-momentum projected wave functions, generated from the
self-consistent solutions of axially symmetric constrained RMF+BCS equations for
$^{154}$Sm.
\subsection{GCM mixing of angular-momentum projected states}
The generator coordinate method (GCM) is based on the assumption that, starting
from a set of intrinsic symmetry-breaking states $\ket{\phi (q)}$ which depend on a collective
coordinate $q$, one can build approximate eigenstates of the nuclear Hamiltonian \cite{RS.80}
\begin{equation}
\ket{\Psi_\alpha} = \sum_j{f_\alpha(q_j)\ket{\phi (q_j)}}\;.
\label{GCM-state}
\end{equation}
Here the basis states $\ket{\phi (q)}$ are Slater
determinants of single-nucleon states generated by solving the
constrained relativistic mean-field + BCS equations with
the mass quadrupole moment as the generating coordinate $q$.
The axially deformed mean-field breaks rotational
symmetry, so that the basis states $\ket{\phi (q)}$ are not eigenstates of the
total angular momentum. To be able to compare
theoretical results with data, it is necessary to construct states
with good angular momentum
\begin{equation}
\ket{\Psi_\alpha^{JM}} = \sum_{j,K}{f_\alpha^{JK}(q_j)\hat{P}_{MK}^J
    \ket{\phi (q_j)}}\;,
\label{AMPGCM-state}
\end{equation}
where $\hat{P}^J_{MK}$ denotes the angular momentum projection operator
\begin{equation}
\hat{P}^J_{MK} = \frac{2J+1}{8\pi^2}\int{d\Omega D_{MK}^{J*}(\Omega )\hat{R}
   (\Omega )}\;.
\label{proj_op}
\end{equation}
The integration is performed over the three Euler angles
$\alpha$, $\beta$, and $\gamma$.
$D_{MK}^{J}(\Omega )=e^{-iM\alpha}d^J_{MK}(\beta)e^{-iK\gamma}$ is the Wigner
function, and
$\hat{R}(\Omega )= e^{-i\alpha\hat{J}_z}$ $e^{-i\beta\hat{J}_y}
e^{-i\gamma\hat{J}_z}$ is the rotation operator.
The weight functions $f_{\alpha}^{JK}(q_j)$ are determined from the variation:
\begin{equation}
 \delta E^{J} =
 \delta \frac{\bra{\Psi_\alpha^{JM}} \hat{H} \ket{\Psi_\alpha^{JM}}}
            {\bra{\Psi_\alpha^{JM}}\Psi_\alpha^{JM}\rangle} = 0 \; ,
\label{variational}
\end{equation}
that is, by requiring that the expectation value of the energy is stationary
with respect to an arbitrary variation $\delta f_{\alpha}^{JK}$. This
leads to the Hill-Wheeler equation
\begin{equation}
\sum_{j,K}f_{\alpha}^{JK}(q_j)
  \left( \left\langle\phi(q_i) \right|\hat{H}\hat{P}_{MK}^J\left|
  \phi(q_j)\right\rangle - E^J_\alpha
  \left\langle\phi(q_i) \right|\hat{P}_{MK}^J\left|\phi(q_j)\right\rangle \right) = 0\;.
\label{HWEQ}
\end{equation}
The restriction to
axially symmetric configurations ($\hat{J}_z\ket{\phi(q)} = 0$)
simplifies the problem considerably, because in this case the
integrals over the Euler angles $\alpha$ and $\gamma$ can be
performed analytically, and for an arbitrary multipole operator
$\hat{Q}_{\lambda \mu}$ one thus finds
\begin{eqnarray}
\bra{\phi(q_i)}\hat{Q}_{\lambda \mu} \hat{P}_{MK}^J\ket{\phi(q_j)}=  &
(2J+1)\frac{1+(-1)^J}{2}\delta_{M-\mu}\delta_{K0}
  \int_0^{\pi/2}{d\beta \sin{\beta}~d_{-\mu 0}^{J*}(\beta)} \nonumber \\
&  \bra{\phi(q_i)} \hat{Q}_{\lambda \mu}e^{-i\beta\hat{J}_y}\ket{\phi(q_j)}\;.
\label{matel2}
\end{eqnarray}
This expression vanishes for odd values of angular momentum $J$ and,
therefore, all projected quantities are defined only for even values of $J$.
The norm overlap kernel
\begin{eqnarray}
\mathcal{N}^J(q_i,q_j) &=& \bra{\phi(q_i)} \hat{P}_{MK}^J\ket{\phi(q_j)} =
\nonumber \\
&& (2J+1)\frac{1+(-1)^J}{2}\delta_{M0}\delta_{K0}
  \int_0^{\pi/2}{d\beta \sin{\beta}~d_{00}^{J*}(\beta) }\nonumber \\
  && \bra{\phi(q_i)}
  e^{-i\beta\hat{J}_y}\ket{\phi(q_j)}\;,
\label{normker}
\end{eqnarray}
can be evaluated by employing
the generalized Wick theorem:
\begin{equation}
n(q_i,q_j;\beta) \equiv \bra{\phi(q_i)} e^{-i\beta\hat{J}_y}\ket{\phi(q_j)}
  = \pm\sqrt{det~\mathcal{N}_{ab}(q_i,q_j;\beta)}\;.
\label{norm}
\end{equation}
The overlap matrix is defined:
\begin{equation}
\mathcal{N}_{ab}(q_i,q_j;\beta) = u_a(q_i)R_{ab}(q_i,q_j;\beta)u_b(q_j) +
                         v_a(q_i)R_{ab}(q_i,q_j;\beta)v_b(q_j) \;,
\label{N_mat}
\end{equation}
where $u$ and $v$ denote the BCS occupation probabilities, the matrix $R$
reads
\begin{equation}
R_{ab}(q_i,q_j;\beta) = \int{\psi_a^\dagger(\bm{r};q_i)e^{-i\beta\hat{J}_y}
                     \psi_b(\bm{r};q_j)d\bm{r}}\;,
\label{R_mat}
\end{equation}
and $\psi_a(\bm{r};q_i)$ denotes the self-consistent intrinsic single-nucleon spinor
at the generating coordinate $q_i$. The Hamiltonian kernel
\begin{eqnarray}
\mathcal{H}^J(q_i,q_j) &=& \bra{\phi(q_i)}\hat{H} \hat{P}_{MK}^J\ket{\phi(q_j)} =
\nonumber \\ &&
(2J+1)\frac{1+(-1)^J}{2}\delta_{M0}\delta_{K0}
  \int_0^{\pi/2}{d\beta \sin{\beta}~d_{00}^{J*}(\beta) } \nonumber \\
 && \bra{\phi(q_i)}\hat{H}
  e^{-i\beta\hat{J}_y}\ket{\phi(q_j)}\;,
\label{hamker}
\end{eqnarray}
can be calculated from the energy functional provided the
{\em intrinsic} densities are replaced by {\em transition}
one-body matrices when evaluating the expression
\begin{equation}
h(q_i,q_j;\beta)\equiv \bra{\phi(q_i)}\hat{H} e^{-i\beta\hat{J}_y}\ket{\phi(q_j)}
  =\int{d\bm{r}~\mathcal{E}_{tot}(\bm{r};q_i,q_j,\beta)}\;.
\end{equation}

The basis states $\ket{\phi(q_j)}$ are not eigenstates of the proton
and neutron number operators $\hat{Z}$ and $\hat{N}$. The adjustment of the
Fermi energies in a BCS calculation ensures only that the average value of
the nucleon number operators corresponds to the actual number of nucleons.
Consequently, the wave functions $\ket{\Psi_\alpha^{JM}}$ are generally not
eigenstates of the nucleon number operators and, moreover, the average values
of the nucleon number operators are not necessarily equal to the number of
nucleons in a given nucleus. To restore the
correct mean values of the nucleon numbers, one can
modify the Hill-Wheeler equation
by replacing $h(q_i,q_j;\beta)$ with
\begin{equation}
h^\prime(q_i,q_j;\beta) = h(q_i,q_j;\beta)
 -\lambda_p\left[ z(q_i,q_j;\beta)-z_0\right]
  -\lambda_n\left[ n(q_i,q_j;\beta)-n_0\right]\;,
\end{equation}
where
\begin{equation}
z(q_i,q_j;\beta)=\bra{\phi(q_i)}\hat{Z} e^{-i\beta\hat{J}_y}\ket{\phi(q_j)}\;,
\end{equation}
and
\begin{equation}
n(q_i,q_j;\beta)=\bra{\phi(q_i)}\hat{N} e^{-i\beta\hat{J}_y}\ket{\phi(q_j)}\;.
\end{equation}
$\lambda_{p(n)}$ is the proton (neutron) Fermi energy, while $z_0$ and
$n_0$ denote the desired number of protons and neutrons, respectively.

The Hill-Wheeler equation
\begin{equation}
\sum_j{\mathcal{H}^J(q_i,q_j)f^J_\alpha(q_j)} = E^J_\alpha
        \sum_j{\mathcal{N}^J(q_i,q_j)f^J_\alpha(q_j)}\;,
\label{HWEQ2}	
\end{equation}
presents a generalized eigenvalue problem, and thus the weight functions
$f^J_\alpha(q_i)$ are not orthogonal and cannot be interpreted as collective
wave functions for the variable $q$. It is useful to re-express
Eq. (\ref{HWEQ2}) in terms of another set of functions,  $g^J_\alpha(q_i)$,
defined by
\begin{equation}
g^J_\alpha(q_i) = \sum_j(\mathcal{N}^{J})^{1/2}(q_i,q_j) f^J_\alpha(q_j)\;.
\label{coll_wf}
\end{equation}
With this transformation the
Hill-Wheeler equation defines an ordinary eigenvalue problem
\begin{equation}
\sum_j{\tilde{\mathcal{H}}^J(q_i,q_j)g^J_\alpha(q_j)} =E_\alpha g_\alpha^J(q_i)\;,
\end{equation}
with
\begin{equation}
\tilde{\mathcal{H}}^J(q_i,q_j) = \sum_{k,l}(\mathcal{N}^{J})^{-1/2}(q_i,q_k)
  \mathcal{H}^J(q_k,q_l) (\mathcal{N}^{J})^{-1/2}(q_l,q_j) \;.
\end{equation}
The functions $g^J_\alpha(q_i)$ are orthonormal and play the role
of collective wave functions.

Once the weight functions $f^J_\alpha(q)$ are known, it is
straightforward to calculate all physical observables, such as transition
probabilities and spectroscopic quadrupole moments.
The reduced transition probability for a transition between an initial state
$(J_i,\alpha_i)$, and a final state $(J_f,\alpha_f)$, reads
\begin{equation}
B(E2; J_i\alpha_i \to J_f\alpha_f) = \frac{e^2}{2J_i+1}\left| \sum_{q_f,q_i}{
f^{J_f*}_{\alpha_f}(q_f)\bra{J_fq_f}|\hat{Q}_2|\ket{J_iq_i}
f^{J_i}_{\alpha_i}(q_i)}\right|^2\;.
\label{BE2-GCM}
\end{equation}
\subsection{Angular momentum projection and
configuration mixing: $^{154}$Sm}
\begin{figure}
\begin{center}
\vspace{-2cm}
\includegraphics[scale=0.5]{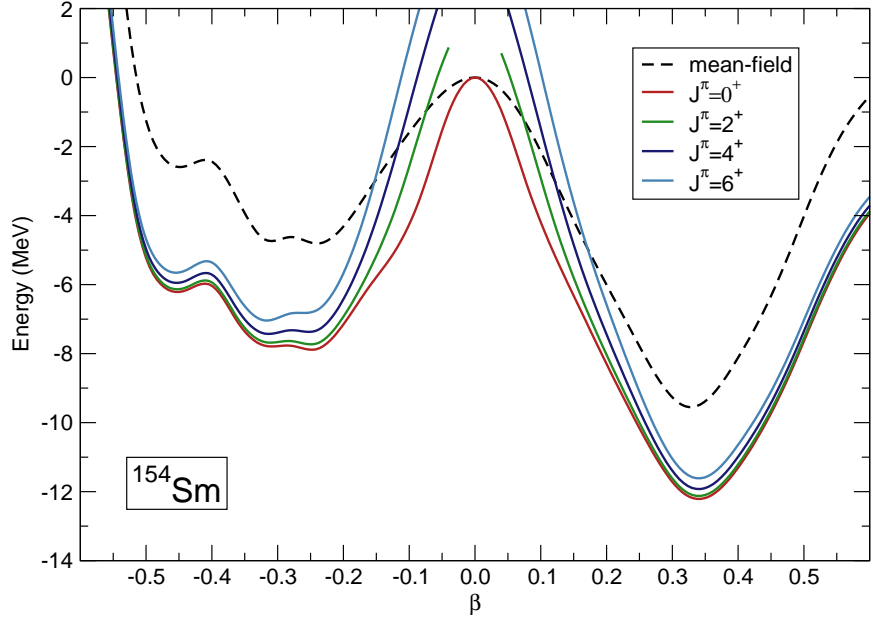}
\end{center}
\caption{\label{PEC-154sm} The RMF+BCS binding energy curve of $^{154}$Sm
(dashed), and the corresponding angular-momentum projected
($J^\pi=0^+ ,2^+ ,4^+$ , and $6^+$) energy curves, as functions of the axial
deformation $\beta$.}
\end{figure}

As an illustration of GCM configuration mixing of angular-momentum projected
states, we consider the case of the prolate deformed, axially-symmetric ground
state of $^{154}$Sm.
The intrinsic symmetry-breaking wave functions that are used in the configuration
mixing calculation are obtained as solutions of the self-consistent
relativistic mean-field equations, subject to constraint on the
mass quadrupole moment. The interaction in the particle-hole
channel is determined by the density functional DD-PC1,
and a density-independent $\delta$-force is
used as the effective interaction in the particle-particle channel,
supplemented with a smooth cut-off determined by a Fermi
function in the single-particle energies \cite{BMM.02}.
Pairing correlations are treated within the BCS framework, and
the pairing contribution to the total energy is given by
\begin{equation}
{E}_{pair}^{p(n)} = \int{\mathcal{E}_{pair}^{p(n)}(\bm{r})d\bm{r}}=
   \frac{V_{p(n)}}{4}\int{\kappa_{p(n)}^*(\bm{r})\kappa_{p(n)}(\bm{r}) d\bm{r}}\;,
\label{epair}	
\end{equation}
for protons and neutrons, respectively. $\kappa_{p(n)}(\bm{r})$ denotes
the local part of the pairing tensor, and $V_{p} = -321$ MeV fm$^3$ and
$V_{n} = -308$ MeV fm$^3$ are the pairing strength
parameters adjusted to empirical pairing gaps. 
We note that while the BCS constant-gap approximation was employed in the 
fit of the parameters of the functional DD-PC1 to experimental binding energies 
(cf. Sec.~\ref{DD-PC1}), obviously constant gaps cannot be used in a constrained 
calculation. Constant gaps that are adjusted to empirical values correspond only to 
the ground state and, therefore, cannot be used to calculate configurations that 
generally represent excited states. This is the reason for employing a zero-range 
pairing force in this example of angular momentum projection and
configuration mixing calculation.

The GCM basis is constructed from the self-consistent solution of
constrained single-nucleon Dirac equations on a regular mesh in the
generating coordinate -- the mass quadrupole moment. The corresponding
axial deformation parameter spans the interval from $\beta=-0.68$
to $\beta=1$, with a spacing $\Delta \beta=0.04$. The GCM basis
thus consists of 43 intrinsic states. The large and small
components of the Dirac spinors are expanded in a basis of axially symmetric
oscillator eigenfunctions for $N=14$ major shells.

\begin{figure}
\vspace{-1cm}
\begin{center}
\includegraphics[scale=0.525]{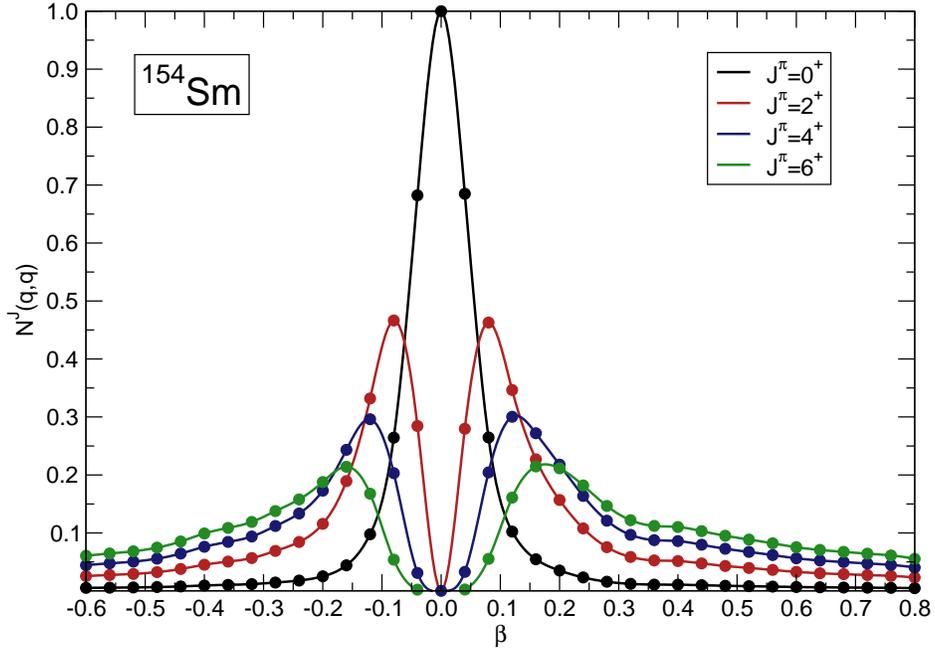}
\end{center}
\caption{\label{norm_J} Projected norm overlap kernels $\mathcal{N}^J(q,q)$
as a functions of the axial quadrupole deformation $\beta$.}
\end{figure}

The resulting binding energy curve of $^{154}$Sm is shown in Fig.~\ref{PEC-154sm},
as function of the axial deformation $\beta$.
The RMF+BCS mean-field calculation yields a prolate ground-state
minimum at $\beta = 0.32$, more than 4.5 MeV deeper than the wide local minimum on
the oblate side. One might notice that for prolate shapes
the potential is rather stiff with respect to $\beta$ around the minimum,
and this means that the structures built on the ground state will be localized in
this minimum, that is, there will be very little mixing with oblate structures for
states with relatively low angular momenta. Fig.~\ref{PEC-154sm} also displays the
corresponding angular-momentum projected ($J^\pi=0^+ ,2^+ ,4^+$, and $6^+$) energy curves.
At this stage we do not consider configuration mixing yet,
and the projected energy of the $\ket{\phi(q)}$ state reads
\begin{equation}
E^J(q)=\frac{\mathcal{H}^J(q,q)}{\mathcal{N}^J(q,q)}\;,
\label{eproj}
\end{equation}
where the projected norm overlap kernels
\begin{equation}
\mathcal{N}^J(q,q)=\bra{\phi(q)}P^J_{00}\ket{\phi(q)}
\end{equation}
are plotted in Fig.~\ref{norm_J}. The spherical configuration is a pure $0^+$ state
($\mathcal{N}^{J=0}(0,$ $0)= 1$), and the maxima of the projected norm overlap
kernels for higher angular momenta are correspondingly shifted to larger
deformations.

Coming back to the projected energy curves in Fig.~\ref{PEC-154sm},
we notice that, since the spherical configuration
is already a pure $0^+$ state, there is no energy gain for $J^\pi = 0^+$
at $\beta =0$. The rotational energy correction at the prolate minimum
$\beta=0.32$ is more than 2.5 MeV, and even larger in the region of
the oblate minimum. The spherical point $\beta =0$ is not included in the plots
of $E^J(q)$ for $J\ge2$. Namely, for $J \neq 0$ the quantities
$\mathcal{H}^J(0,0)$ and $\mathcal{N}^J(0,0)$ are so small, that
their ratio in Eq.~(\ref{eproj}) cannot be determined accurately.
For higher values of the angular momentum several additional configurations
close to the spherical point are also characterized by very small values
of the projected norm overlap kernel. These configurations can safely be
omitted from the projected energy curves, because on the one hand
the angular momentum projection becomes inaccurate at these points,
and on the other hand the corresponding angular momentum projected
states would not play any role in configuration mixing calculations.

The next step in the solution of the Hill-Wheeler equation (\ref{HWEQ2}) is the
diagonalization of the norm overlap kernel $\mathcal{N}^J(q_i,q_j)$
\begin{equation}
\sum_j{\mathcal{N}^J(q_i,q_j)u_k(q_j)} = n_ku_k(q_i)\;.
\label{diag_norm}
\end{equation}
Since the basis functions
$\ket{\phi (q_i)}$ are not linearly independent, many
eigenvalues $n_k$ will be very close to zero. The corresponding eigenfunctions
$u_k(q_i)$ are rapidly oscillating and carry very little physical information.
However, due to numerical uncertainties, their contribution to
$\tilde{\mathcal{H}}^J(q_i,q_j)$ can be large, and these states should be
removed from the basis. From the remaining states one builds the collective
Hamiltonian
\begin{equation}
\mathcal{H}^{Jc}_{kl} = \frac{1}{\sqrt{n_k}}\frac{1}{\sqrt{n_l}}
  \sum_{i,j}{u_k(q_i)\tilde{\mathcal{H}}^J(q_i,q_j)u_l(q_j)}\;,
\label{Hcoll}
\end{equation}
which is subsequently diagonalized
\begin{equation}
\sum_{k,l}\mathcal{H}^{Jc}_{kl}g_l^{J\alpha} = E^J_{\alpha}g_k^{J\alpha} \;.
\label{mat_coll}
\end{equation}
The solution determines both the ground state energy, and the energies
of excited states, for each value of the angular momentum $J$.
The collective wave functions $g_\alpha^J(q)$, and the weight functions
$f_\alpha^J(q)$, are calculated from the norm overlap eigenfunctions
\begin{equation}
g_\alpha^J(q_i) = \sum_l{ g_l^{J\alpha} u_l(q_i) }\;,
\label{g_u}
\end{equation}
and
\begin{equation}
f_\alpha^J(q_i) = \sum_l{\frac{g_l^{J\alpha}}{\sqrt{n_l}} u_l(q_i)} \;.
\label{f_u}
\end{equation}
\begin{figure}
\vspace{-1cm}
\begin{center}
\includegraphics[scale=0.525]{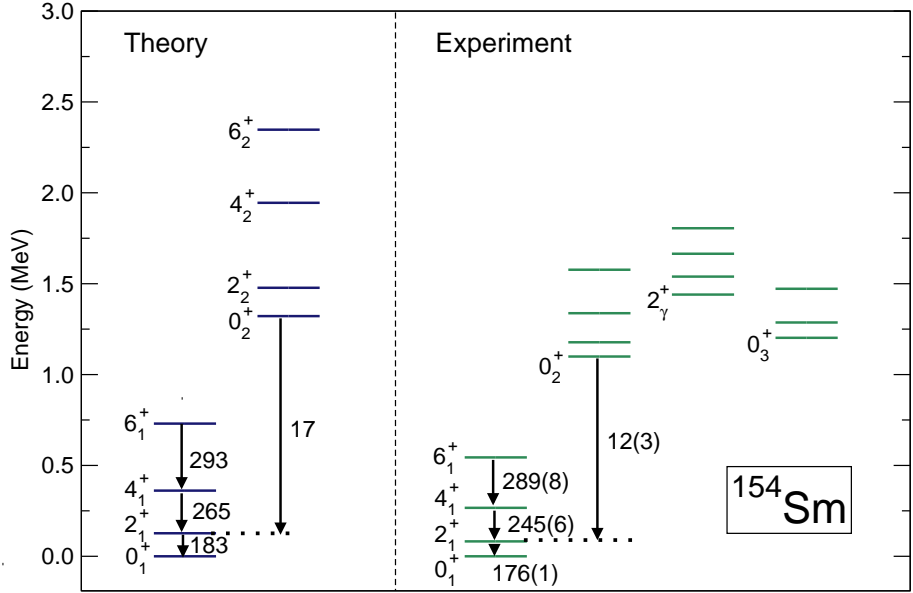}
\end{center}
\caption{\label{154sm-spectrum} Angular-momentum projected GCM results for the
excitation energies and B(E2) values (in Weisskopf units) of the
lowest two bands in $^{154}$Sm, in comparison with available data. }
\end{figure}

The GCM excitation energies and the corresponding B(E2) values for
the two lowest bands in $^{154}$Sm: the ground-state band and the $\beta$-band,
are displayed in Fig.~\ref{154sm-spectrum}. The results of the angular-momentum
projected (AMP) configuration mixing calculation are compared with available data.
Since the present calculation is restricted to axial symmetry, it cannot describe structures
based on the $\gamma$ degree of freedom. Considering that, for a given EDF and
the effective pairing interaction, the AMP+GCM calculation is parameter-free, the
agreement with data is remarkable. One might notice the excellent agreement
of the transition probabilities, calculated with the bare proton charge, with the experimental
values for transitions within the ground-state band, and the transition from the
band-head of the $\beta$-band to the ground-state band. The calculated excitation
energy of the $\beta$ band-head is above the experimental $0^+_2$ , indicating that
the potential is probably too stiff with respect to $\beta$, and the corresponding moment
of inertia appears to be considerably lower than the effective empirical value. The
difference between the calculated and empirical moments of inertia is much less
pronounced in the ground-state band.

\begin{figure}
\vspace{-1cm}
\begin{center}
\includegraphics[scale=0.475]{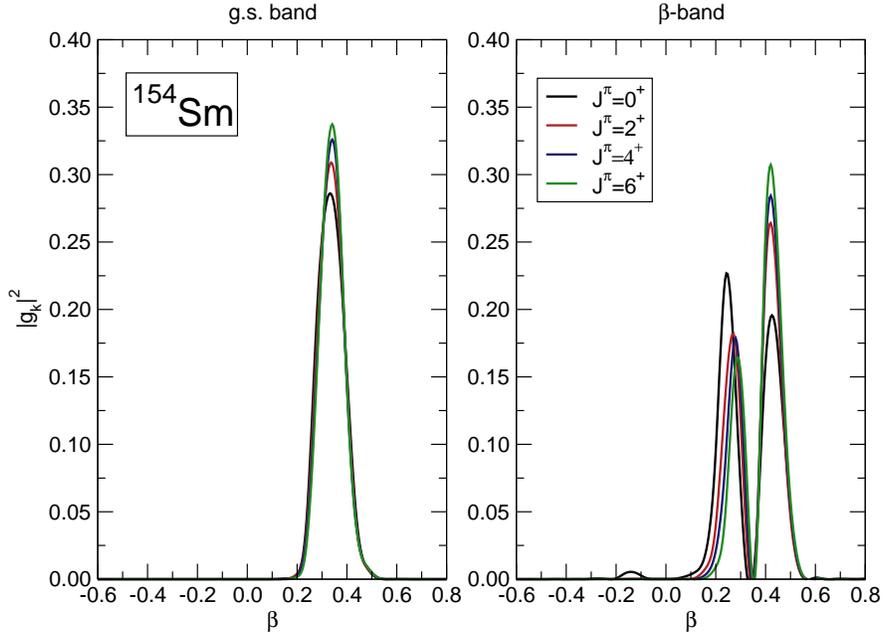}
\end{center}
\caption{\label{154sm-wf} Amplitudes of the angular-momentum projected GCM
collective wave functions $|g_{\alpha}^J(\beta)|^2$ ($J^\pi=0^+ ,2^+ ,4^+$, and $6^+$)
for the ground-state band (left) and the $\beta$-band (right) in $^{154}$Sm. }
\end{figure}

In Fig.~\ref{154sm-wf} we plot the amplitudes of the collective wave
functions $|g_{\alpha}^J(\beta)|^2$
Eq.~(\ref{g_u}) for the two lowest GCM states of each angular momentum
$J^\pi=0^+ ,2^+ ,4^+$, and $6^+$. The amplitudes in the left panel correspond to
the states of the ground-state band, whereas the collective functions of the $\beta$-band
are plotted in the right panel. For both bands the collective
functions are localized in the prolate well, and only for the state $0^+_2$ the
wave function displays a small amount of oblate admixture.

The treatment of collective correlations in REDF-based structure models has been
extended in Ref.~\cite{NVR.06b} to axially symmetric GCM configuration
mixing of angular-momentum and particle-number projected states, and very
recently a model has been developed that includes triaxial angular-momentum
projection \cite{Yao.09,Yao.10}. The latter does not involve particle number projection.
It has to be emphasized that, while GCM configuration mixing of axially symmetric states
has been implemented by several groups and routinely used in nuclear
structure studies, the application of this method to triaxial shapes presents
a much more involved and technically difficult problem. Only the most
recent advances in parallel computing and modeling have enabled
the implementation of models \cite{BH.08,RE.10}, based on triaxial
symmetry-breaking intrinsic states, that are projected on particle number
and angular momentum, and finally mixed by the generator coordinate method.
This implementation is equivalent to a seven-dimensional GCM calculation,
mixing all five degrees of freedom of the quadrupole operator and the
gauge angles for protons and neutrons. The numerical
realization, however, is very complex, and applications to
medium-heavy and heavy nuclei are still computationally too demanding
and time-consuming. This is true even for a model that does not include
particle number projection \cite{Yao.10}. In addition, the use of general EDFs
with an arbitrary dependence on nucleon densities in GCM-type
calculations, often leads to discontinuities or even divergences of
the energy kernels as a function of deformation~\cite{AER.01,Dob.07}.
Only for a specific type of density dependence a regularization method
can be implemented~\cite{LDB.09,BDL.09}, that corrects the energy kernels and
removes the discontinuities and divergences. In the next section we therefore
review an approximation to the full GCM approach, based on the microscopic
REDF framework, that includes rotational symmetry restoration and takes
into account triaxial quadrupole fluctuations in arbitrary heavy nuclei.

\section{Collective Hamiltonian in five dimensions based on relativistic EDFs}
\label{H_coll}
In an alternative approach to five-dimensional quadrupole dynamics
that restores rotational symmetry and allows for
fluctuations around the triaxial mean-field minima, a collective Bohr Hamiltonian
can be formulated, with deformation-dependent parameters determined by
microscopic self-consistent mean-field calculations. There are two
principal approaches to derive the collective Hamiltonian starting from
a microscopic framework based on an effective inter-nucleon interaction
or energy density functional: (i) the adiabatic approximation to the
time-dependent HFB theory (ATDHFB) \cite{BV.78}, and (ii) the generator
coordinate method (GCM) with the Gaussian overlap approximation
(GOA) \cite{RG.87,Bon.90,PR.09}.
With the assumption that the GCM overlap kernels can be
approximated by Gaussian functions \cite{RS.80}, the local expansion of the
kernels up to second order in the non-locality transforms the
GCM Hill-Wheeler equation into a second-order differential equation -
the Schr\"odinger equation for the collective Hamiltonian. The
kinetic part of this Hamiltonian contains an inertia
tensor \cite{GG.75}, and the potential energy is determined by the
diagonal elements of the Hamiltonian kernel, and also includes zero-point
energy (ZPE) corrections \cite{GG.79}.

The dynamics of the collective Bohr Hamiltonian is determined by the
vibrational inertial functions and the moments of inertia \cite{GR.80}.
For these quantities either the GCM-GOA (Yoccoz masses~\cite{PY.57})
or the ATDHFB expressions  (Thouless-Valatin masses~\cite{TV.62})
can be used. The Thouless-Valatin masses
have the advantage that they also include the time-odd components of
the mean-field potential and, in this sense, the full dynamics of a
nuclear system. In the GCM approach these components can only be included if,
in addition to the coordinates  $q_i$, the corresponding
canonically conjugate momenta $p_i$ are also taken into account, but this
is obviously a very complicated task. In many applications a further
simplification is thus introduced in terms of cranking formulas \cite{Ing.56,GG.79},
that represent the perturbative limit for the Thouless-Valatin masses, and the
corresponding expressions for ZPE corrections. This approximation was
applied in recent studies using models based both on the Gogny
interaction \cite{LGD.99,Cle.07}, and Skyrme energy density functionals \cite{Pro.04,PR.09}.
The approximate inclusion of Thouless-Valatin corrections to the
mass parameters and moments of inertia of the Skyrme-based Bohr Hamiltonian
was discussed in Ref.~\cite{Pro.04}.
In a recent systematic study \cite{DGL.10} of low-energy nuclear structure at
normal deformation,  based on the Hartree-Fock-Bogoliubov theory extended by
the generator coordinate method and mapped onto a five-dimensional
collective quadrupole Hamiltonian, the Thouless-Valatin moments of inertia were used,
whereas the cranking approximation was used for the quadrupole mass parameters.
Using the Gogny D1S interaction, even-even nuclei with proton numbers $Z=10$ to
$Z=110$ and neutron numbers $N\leq 200$ were calculated.

Here we review a recent implementation for the solution of the eigenvalue problem
of a five-dimensional collective Hamiltonian for quadrupole vibrational and rotational
degrees of freedom, with parameters determined by constrained self-consistent
relativistic Hartree-Bogoliubov calculations for triaxial shapes
\cite{PR.04,Ni.09,Li.09a,Li.09b,Li.10a,Li.10b}.
\subsection{\label{secVb}Collective Hamiltonian}
The general Bohr collective model for the description of quadrupole
collective states, including a detailed discussion of the model's kinematics,
has recently been reviewed in Ref.~\cite{PR.09}.
Nuclear excitations determined by quadrupole vibrational and
rotational degrees of freedom can be treated simultaneously by
considering five quadrupole collective coordinates
$\alpha_\mu,\;\mu =-2,-1,\dots,2$ that describe the surface of
a deformed nucleus: $R=R_0[1+\sum_{\mu}\alpha_\mu Y^*_{2\mu}]$.
To separate rotational and vibrational motion,
these coordinates are usually parameterized in terms of two
deformation parameters $\beta$ and $\gamma$, and three Euler angles
$(\phi,\;\theta,\;\psi)\equiv \Omega$ that define the orientation
of the intrinsic principal axes in the laboratory frame
\begin{equation}
\alpha_\mu = D_{\mu 0}^2(\Omega)\beta\cos{\gamma}+\frac{1}{\sqrt{2}}
   \left[D_{\mu  2}^2(\Omega)+D_{\mu -2}^2(\Omega)\right]\beta \sin{\gamma}\; ,
\end{equation}
where $D^\lambda_{\mu \nu}$ is the Wigner function. The
three terms of the classical collective Hamiltonian, expressed in
terms of the intrinsic variables $\beta$, $\gamma$ and Euler angles
\begin{equation}
\label{hamiltonian-cl}
H_{\textnormal{coll}} = \mathcal{T}_{\textnormal{vib}}(\beta,\gamma)
                            +\mathcal{T}_{\textnormal{rot}}(\beta,\gamma,\Omega)
                            +\mathcal{V}_{\textnormal{coll}}(\beta,\gamma)\; ,
\end{equation}
denote the contributions from the vibrational kinetic energy:
\begin{equation}
\mathcal{T}_{\textnormal{vib}} = \frac{1}{2}B_{\beta\beta}\dot{\beta}^2
   + \beta B_{\beta\gamma} \dot{\beta}\dot{\gamma}
   +\frac{1}{2} \beta^2B_{\gamma\gamma}\dot{\gamma}^2\; ,
\end{equation}
the rotational kinetic energy:
\begin{equation}
\mathcal{T}_{\textnormal{rot}} = \frac{1}{2}\sum_{k=1}^3{\mathcal{I}_k\omega_k^2},
\end{equation}
and the collective potential energy
$\mathcal{V}_{\textnormal{coll}}(\beta,\gamma)$. The mass parameters
$B_{\beta\beta}$, $B_{\beta\gamma}$, $B_{\gamma\gamma}$, and the
moments of inertia $\mathcal{I}_k$ depend on the quadrupole
deformation variables $\beta$ and $\gamma$.

After quantization the classical kinetic energy of the Hamiltonian
Eq.~(\ref{hamiltonian-cl}):
\begin{equation}
T=\frac{1}{2}\sum_{ij}{B_{ij}(q)\dot{q}_i\dot{q}_j} \; ,
\end{equation}
reads:
\begin{equation}
\hat{H}_{\textnormal{kin}} = -\frac{\hbar^2}{2}\frac{1}{\sqrt{\textnormal{det} B}}
                          \sum_{ij}{\frac{\partial}{\partial q_i}
         \sqrt{\textnormal{det} B}(B^{-1})_{ij}\frac{\partial}{\partial q_j} }.
\end{equation}
The kinetic energy tensor takes the block diagonal form:
\begin{equation}
B = \left( \begin{array}{cc} B_{\textnormal{vib}} & 0 \\
                                                  0 & B_{\textnormal{rot}} \end{array} \right) \;,
\end{equation}
with the vibrational part of the tensor
\begin{equation}
B_{\textnormal{vib}} = \left( \begin{array}{cc} B_{\beta\beta} & \beta B_{\beta\gamma} \\
                                          \beta B_{\beta\gamma} &\beta^2 B_{\gamma \gamma}
        \end{array} \right).
\end{equation}
In general, the rotational part is a complicated function of the Euler
angles but, using the quasi-coordinates related to the components of
the angular momentum in the body-fixed frame, it takes a simple
diagonal form
\begin{equation}
\left(B_{\textnormal{rot}}\right)_{ik}=\delta_{ik}\mathcal{I}_k, \quad k=1,2,3 \;,
\end{equation}
with the moments of inertia expressed as
\begin{equation}
\mathcal{I}_k = 4B_k\beta^2\sin^2(\gamma-2k\pi/3) \;.
\end{equation}
This particular functional form is motivated by the fact that
all three moments of inertia vanish for the spherical configuration
($\beta=0$) and, additionally, $\mathcal{I}_z$ and $\mathcal{I}_y$ vanish for axially
symmetric prolate ($\gamma=0^0$) and oblate ($\gamma=60^0$)
configurations, respectively.
The  resulting determinant reads
\begin{equation}
\label{detB}
\textnormal{det} B = \textnormal{det} B_{\textnormal{vib}}\cdot
                                \textnormal{det} B_{\textnormal{rot}}
  = 4 wr\beta^8 \sin^2{3\gamma} \; ,
\end{equation}
where $w=B_{\beta\beta}B_{\gamma\gamma}-B_{\beta\gamma}^2 $ and
   $r=B_1B_2B_3$.
The quantized collective Hamiltonian can hence be written in the form:
\begin{equation}
\label{hamiltonian-quant}
\hat{H} = \hat{T}_{\textnormal{vib}}+\hat{T}_{\textnormal{rot}}
              +V_{\textnormal{coll}} \; ,
\end{equation}
with
\begin{align}
\hat{T}_{\textnormal{vib}} =&-\frac{\hbar^2}{2\sqrt{wr}}
   \left\{\frac{1}{\beta^4}
   \left[\frac{\partial}{\partial\beta}\sqrt{\frac{r}{w}}\beta^4
   B_{\gamma\gamma} \frac{\partial}{\partial\beta}
   - \frac{\partial}{\partial\beta}\sqrt{\frac{r}{w}}\beta^3
   B_{\beta\gamma}\frac{\partial}{\partial\gamma}
   \right]\right.
   \nonumber \\
   &+\frac{1}{\beta\sin{3\gamma}}\left.\left[
   -\frac{\partial}{\partial\gamma} \sqrt{\frac{r}{w}}\sin{3\gamma}
      B_{\beta \gamma}\frac{\partial}{\partial\beta}
    +\frac{1}{\beta}\frac{\partial}{\partial\gamma} \sqrt{\frac{r}{w}}\sin{3\gamma}
      B_{\beta \beta}\frac{\partial}{\partial\gamma}
   \right]\right\} \; ,
\end{align}
and
\begin{equation}
\hat{T}_{\textnormal{\textnormal{\textnormal{rot}}}} =
\frac{1}{2}\sum_{k=1}^3{\frac{\hat{J}^2_k}{\mathcal{I}_k}} \; ,
\end{equation}
where $\hat{J}_k$ denotes the components of the angular momentum in
the body-fixed frame of a nucleus. $V_{\textnormal{coll}}$ is the collective
potential. The Hamiltonian describes quadrupole vibrations,
rotations, and the coupling of these collective modes. The
determinant Eq.~(\ref{detB}) determines the volume element in the
collective space:
\begin{equation}
\label{measure}
\int{d\tau_{coll}}=\int{d\Omega d\tau_0\sqrt{wr}}=
\int_0^\infty{d\beta \beta^4 \int_0^{2\pi}{d\gamma|\sin{3\gamma}|
         \int{d\Omega \sqrt{wr}}}} \; ,
\end{equation}
and the quantized Hamiltonian Eq.~(\ref{hamiltonian-quant}) is
hermitian with respect to the collective measure Eq.~(\ref{measure}).

The eigenvalue problem of the general collective Hamiltonian
Eq.~(\ref{hamiltonian-quant}) can be solved by a direct numerical solution
of a system of partial differential equations using finite-difference
methods, or by employing an expansion of eigenfunctions in terms of a
truncated basis in the collective Hilbert space. The basis functions
depend on the deformation variables
$\beta$ and $\gamma$, and the Euler angles $\phi$, $\theta$ and $\psi$.
In the latter case the eigenvalue problem reduces
to a simple matrix diagonalization, and the main task is the
construction of an appropriate basis for each value of the angular
momentum quantum number. In the implementation developed in
Ref.~\cite{Ni.09} we employed the basis expansion approach.

The diagonalization of the Hamiltonian yields the excitation energies and
collective wave functions:
\begin{equation}
\label{wave-coll}
\Psi_\alpha^{JM}(\beta,\gamma,\Omega) =
  \sum_{K\in \Delta J}
           {\psi_{\alpha K}^J(\beta,\gamma)\Phi_{MK}^J(\Omega)}.
\end{equation}
The angular part corresponds to a linear combination of Wigner
functions
\begin{equation}
\label{Wigner}
\Phi_{MK}^J(\Omega)=\sqrt{\frac{2J+1}{16\pi^2(1+\delta_{K0})}}
\left[D_{MK}^{J*}(\Omega)+(-1)^JD_{M-K}^{J*}(\Omega) \right] \; ,
\end{equation}
and the summation in Eq. (\ref{wave-coll}) is over the allowed set  of
the $K$ values:
\begin{equation}
\Delta J = \left\{ \begin{array}{lcl}
   0,2,\dots,J \quad &\textnormal{for} \quad  &J\; \textnormal{mod}\; 2 = 0 \\
   2,4,\dots,J-1 \quad &\textnormal{for} \quad   &J\; \textnormal{mod}\; 2 =1\; .
\end{array} \right .
\end{equation}
Using the collective wave functions Eq.~(\ref{wave-coll}),
various observables can be calculated and compared with
experimental results. For instance, the quadrupole E2 reduced
transition probability:
\begin{equation}
\label{BE2}
B(\textnormal{E2};\; \alpha J \to \alpha^\prime J^\prime)=
      \frac{1}{2J+1}|\langle \alpha^\prime J^\prime || \mathcal{\hat{M}}(E2) ||
                                        \alpha J  \rangle|^2 \; ,
\end{equation}
where $\mathcal{\hat{M}}(E2)$ is the electric quadrupole operator.
For the $\mathcal{\hat{M}}(E2)$ matrix elements
the current implementation of the model uses a local expression in
the collective deformation variables \cite{KB.67}. This
approximation is justified in the case
of large overlaps between different vibrational amplitudes \cite{LGD.99},
but may be less suited for
transitions between states with a rather small overlap, e.g. for transitions
between super-deformed bands and bands at normal deformation.

The shape of a nucleus can be characterized in a qualitative way
by the average values of the
invariants $\beta^2$, $\beta^3\cos{3\gamma}$, as well as their
combinations. For example, the average value of the invariant
$\beta^2$ in the state $|\alpha J\rangle$:
\begin{equation}
\langle \beta^2\rangle_{J\alpha} = \langle \Psi_\alpha^J | \beta^2 | \Psi_\alpha^J\rangle
=\sum_{K\in\Delta J}{\int{\beta^2|\psi^J_{\alpha,K}(\beta,\gamma)|^2d\tau_0}} \; ,
\end{equation}
and the average values of the deformation parameters
$\beta$ and $\gamma$ in the state
$|\alpha J\rangle$ are calculated from:
\begin{align}
\label{avbeta}
\langle \beta\rangle_{J\alpha} &= \sqrt{\langle \beta^2\rangle_{J\alpha} }, \\
\label{avgamma}
\langle \gamma\rangle_{J\alpha} &=
        \frac{1}{3}\arccos{\frac{\langle \beta^3 \cos{3\gamma}\rangle_{J\alpha}}
            {\sqrt{\langle \beta^2\rangle_{J\alpha} \langle \beta^4\rangle_{J\alpha}}}} ; .
\end{align}
The mixing of different intrinsic configurations in the state
$|\alpha I\rangle$ can be determined from
the distribution of the projection $K$ of the angular momentum $I$
on the $z$ axis in the body-fixed frame:
\begin{equation}
N_K=6\int_0^{\pi/3}{\int_0^\infty{
   |\psi^J_{\alpha,K}(\beta,\gamma)|^2\beta^4|\sin{3\gamma}|d\beta d\gamma}},
\label{NK}
\end{equation}
where the components $\psi^J_{\alpha,K}(\beta,\gamma)$ are defined in
Eq.~(\ref{wave-coll}).
For large deformations the $K$ quantum number is to a good approximation
conserved. Consequently, only one of the integrals Eq.~(\ref{NK})
will give a value close to $1$. A broader distribution of $N_K$
values in the state $|\alpha J\rangle$ provides a measure of mixing
of intrinsic configurations.
\subsection{\label{secVc}Microscopic parameters of the collective Hamiltonian}

The entire dynamics of the collective Hamiltonian is governed by the
seven functions of the intrinsic deformations $\beta$ and $\gamma$:
the collective potential, the three mass parameters:
$B_{\beta\beta}$, $B_{\beta\gamma}$, $B_{\gamma\gamma}$, and the
three moments of inertia $\mathcal{I}_k$. These functions are
determined by the choice of a particular microscopic nuclear energy
density functional or effective interaction.
The entire map of the energy surface as function of the quadrupole
deformation is obtained by imposing constraints on the axial and
triaxial mass quadrupole moments, as described in Sec~\ref{RHB}.
The quasiparticle wave functions and energies, generated from
constrained self-consistent solutions of the RHB model,
provide the microscopic input for the parameters of the
collective Hamiltonian.

In the simplest approximation the moments of inertia are
calculated from the Inglis-Belyaev formula:
\begin{equation}
\label{Inglis-Belyaev}
\mathcal{I}_k = \sum_{i,j}{\frac{| \langle ij |\hat{J}_k | \Phi \rangle |^2}{E_i+E_j}}\quad k=1,2,3,
\end{equation}
where $k$ denotes the axis of rotation, the summation runs over
proton and neutron quasiparticle states
$|ij\rangle=\beta^\dagger_i\beta^\dagger_j|\Phi\rangle$, and
$|\Phi\rangle$ represents the quasiparticle vacuum. The mass
parameters associated with the two quadrupole collective coordinates
$q_0=\langle\hat{Q}_{20}\rangle$ and $q_2=\langle\hat{Q}_{22}\rangle$
are calculated in the cranking approximation:
\begin{equation}
\label{masspar-B}
B_{\mu\nu}(q_0,q_2)=\frac{\hbar^2}{2}
 \left[\mathcal{M}_{(1)}^{-1} \mathcal{M}_{(3)} \mathcal{M}_{(1)}^{-1}\right]_{\mu\nu}\;,
\end{equation}
where
\begin{equation}
\label{masspar-M}
\mathcal{M}_{(n),\mu\nu}(q_0,q_2)=\sum_{i,j}
 {\frac{\left|\langle\Phi|\hat{Q}_{2\mu}|ij\rangle
 \langle ij |\hat{Q}_{2\nu}|\Phi\rangle\right|}
 {(E_i+E_j)^n}}\;,
\end{equation}

The collective energy surface includes the energy of the zero-point
motion, and this quantity has to be subtracted. The collective zero-point
energy (ZPE) corresponds to a superposition of zero-point motion of
individual nucleons in the single-nucleon potential. In the general
case, the ZPE corrections on the potential energy surfaces depend on
the deformation. The ZPE includes terms originating from the vibrational
and rotational kinetic energy, and a contribution of potential energy
\begin{equation}
\Delta V(q_0,q_2)=\Delta V_{\textnormal{vib}}(q_0,q_2)
                            + \Delta V_{\textnormal{rot}}(q_0,q_2)
                            + \Delta V_{\textnormal{pot}}(q_0,q_2) \; .
\end{equation}
The latter is much smaller than the contribution of the kinetic energy,
and is usually neglected~\cite{LGD.99}. Simple prescriptions for the
calculation of vibrational and rotational ZPE have been derived in
Ref.~\cite{GG.79}. Both corrections are calculated in the cranking
approximation, i.e. on the same level of approximation as the mass
parameters and the moments of inertia. The vibrational ZPE is given
by the expression:
\begin{equation}
\label{ZPE-vib}
\Delta V_{\textnormal{vib}}(q_0,q_2) = \frac{1}{4}
\textnormal{Tr}\left[\mathcal{M}_{(3)}^{-1}\mathcal{M}_{(2)}  \right]\;.
\end{equation}
The rotational ZPE is a sum of three terms:
\begin{equation}
\label{ZPE-rot}
\Delta V_{\textnormal{rot}}(q_0,q_2)=\Delta V_{-2-2}(q_0,q_2)+\Delta V_{-1-1}(q_0,q_2)
                                     +\Delta V_{11}(q_0,q_2),
\end{equation}
with
\begin{equation}
\label{ZPE-rotA}
\Delta V_{\mu\nu}(q_0,q_2) = \frac{1}{4}\frac{\mathcal{M}_{(2),\mu\nu}(q_0,q_2)}
        {\mathcal{M}_{(3),\mu\nu}(q_0,q_2)} \; .
\end{equation}

The individual terms are calculated from Eqs.~(\ref{ZPE-rotA}) and (\ref{masspar-M}),
with the intrinsic components of the quadrupole operator defined by:
\begin{equation}
\hat{Q}_{21}=-2iyz \;,\quad \hat{Q}_{2-1}=-2xz\; ,\quad \hat{Q}_{2-2}=2ixy \; .
\end{equation}
The potential $V_{\textnormal{coll}}$ in the collective Hamiltonian
Eq.~(\ref{hamiltonian-quant}) is obtained by subtracting the ZPE
corrections from the total mean-field energy:
\begin{equation}
\label{Vcoll}
{V}_{\textnormal{coll}}(q_0,q_2) = E_{\textnormal{tot}}(q_0,q_2)
  - \Delta V_{\textnormal{vib}}(q_0,q_2) - \Delta V_{\textnormal{rot}}(q_0,q_2) \; .
\end{equation}
\begin{figure}[htb]
\vspace{1cm}
\begin{center}
\includegraphics[scale=0.525]{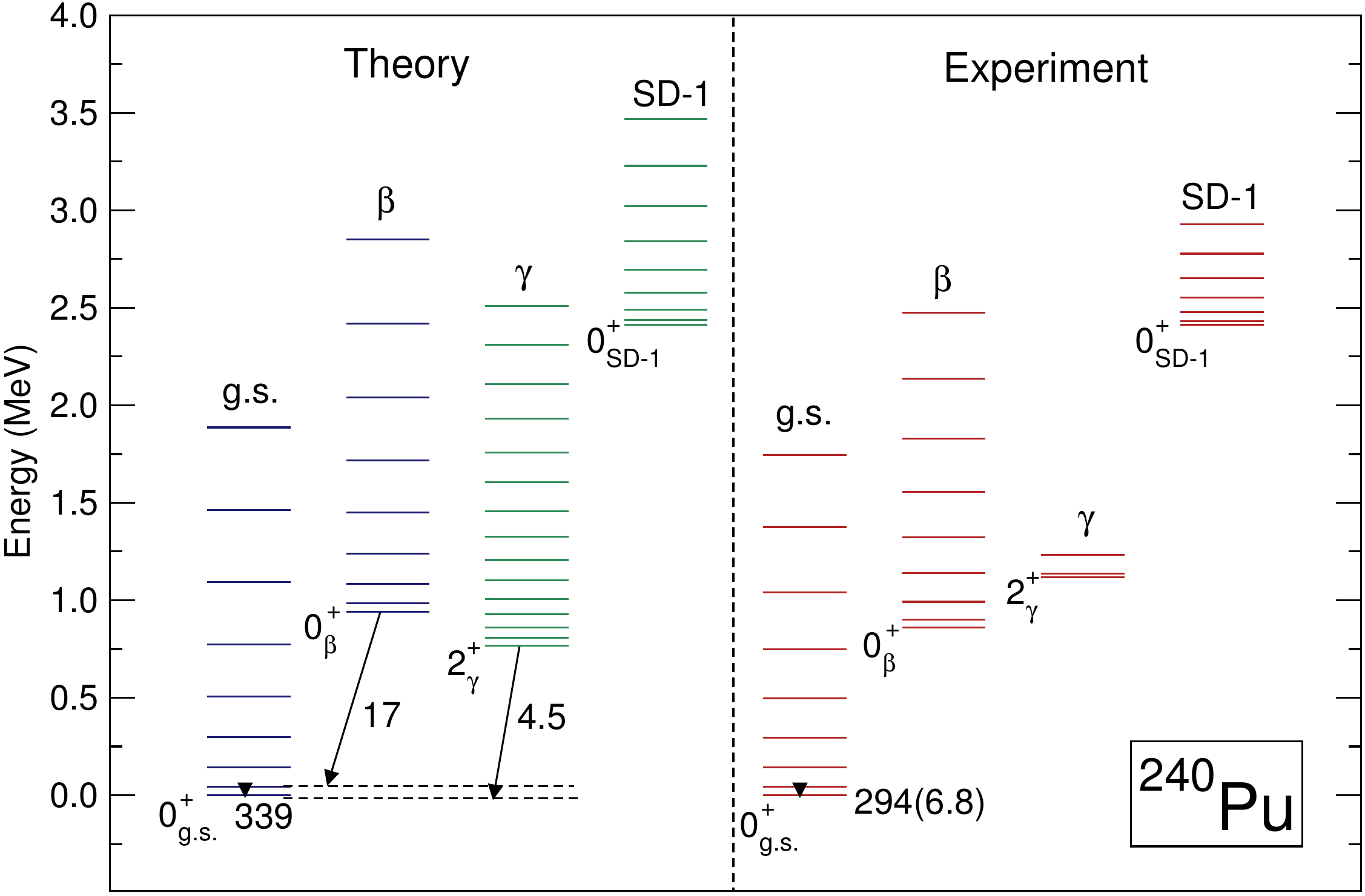}
\end{center}
\caption{\label{spect-240Pu} The low-energy spectrum
of $^{240}$Pu calculated with the DD-PC1 relativistic density functional (left),
compared with data (right) for the three lowest positive-parity bands at normal
deformation, and the lowest $\pi =+$ superdeformed band.}
\end{figure}

As an example in Fig.~\ref{spect-240Pu} we display the spectrum of collective states
of $^{240}$Pu. Starting from constrained self-consistent solutions of the
RHB equations, i.e. using single-quasiparticle energies and wave functions that correspond
to each point on the energy surface shown in Fig.~\ref{PES-240Pu}, the
parameters of the collective Hamiltonian are calculated as functions of
the deformations $\beta$ and  $\gamma$. The diagonalization of the
Hamiltonian yields the excitation spectrum shown in Fig.~\ref{spect-240Pu},
shown in comparison to data for the three lowest positive-parity bands at normal
deformation, and the lowest $\pi =+$ super-deformed band  of $^{240}$Pu.
In addition to the yrast ground-state band, in deformed and transitional nuclei excited
states are also assigned to (quasi) $\beta$ and $\gamma$
bands. This is done according to the distribution of the projection $K$ of the angular
momentum $I$ on the $z$ axis of the body-fixed frame Eq.~(\ref{NK}).
Excited states with predominant $K=2$ components in the
wave function are assigned to the $\gamma$-band, whereas the $\beta$-band
comprises the states above the yrast characterized by dominant $K=0$ components.
$K=0$ states are assigned to the super-deformed band based on the calculated
average value of the deformation parameter $\beta$ Eq.~(\ref{avbeta}).

The Inglis-Belyaev (IB) moments of inertia Eq.~(\ref{Inglis-Belyaev}) of the
collective Hamiltonian have been multiplied by a common factor so that the
calculated energy of the $2_1^+$ state coincides with the experimental value.
This scale parameter reflects the well known fact that the IB expression predicts
effective moments of inertia that are smaller than empirical values. In the
calculation of the spectrum of $^{240}$Pu we have thus followed the prescription
of Ref.~\cite{LGD.99} where, by comparing the more realistic, but also
computationally more involved, Thouless-Valatin (TV) moments of
inertia with the IB values as functions of the axial deformation for
superdeformed bands in the $A \approx 190$ mass region, it was shown that
the TV correction to the perturbative IB expression is
almost independent of deformation, and does not include significant
new structures in the moments of inertia. It was thus suggested that
the moments of inertia to be used in the collective Hamiltonian can
be simply related to the IB values through the minimal prescription:
$\mathcal{I}_k (q) = \mathcal{I}^{IB}_k (q) (1+ \alpha)$, where $q$
denotes the generic deformation parameter, and $\alpha$ is a constant
that can be determined in comparison to data. In the present
case $\alpha=0.32$ for $^{240}$Pu.

When the IB effective moment of inertia is renormalized to the
empirical value, the excitation spectrum of the collective
Hamiltonian determined by the functional DD-PC1 is in very good
agreement with the available data for the ground-state band, $\beta$
and $\gamma$ bands, and even the lowest super-deformed band SD-1.
Compared to the corresponding experimental sequence, the position of
the $\gamma$ band is predicted at somewhat lower excitation energy,
and this might indicate that the theoretical PES is probably too soft
in $\gamma$. The $\beta$-band is calculated at slightly higher energy
compared to experiment, and the predicted position of SD-1 is within
the experimental error bounds. Very few data are available on
electromagnetic transition rates in $^{240}$Pu. In fact, except for
the lifetime of the $2^+_1$ state, only the lifetimes of K-isomers have
been measured but these include configurations not contained in our
collective model space. Therefore, in Fig.~\ref{spect-240Pu} we only
display the calculated B(E2) values, in Weisskopf units (W.u.), for
the transition $2^+_1 \to 0^+_1$ and from the band-heads of the
$\beta$ and $\gamma$ bands to the ground-state band. It should
be emphasized that besides the renormalization of the moment of
inertia, the calculation is completely parameter-free, i.e. by using
structure models based on self-consistent mean-field single-particle
solutions, physical observables, such as transition probabilities and
spectroscopic quadrupole moments, are calculated in the full
configuration space and there is no need for effective charges. Using
the bare value of the proton charge in the electric quadrupole
operator $\mathcal{\hat{M}}(E2)$, the transition probabilities
between eigenstates of the collective Hamiltonian can be directly
compared to data.
\subsection{\label{secVd}Illustrative calculation: evolution of triaxial shapes in
Pt isotopes}
Most deformed nuclei display axially-symmetric prolate ground-state shapes,
but some regions of the nuclide chart are characterized by the
occurrence of oblate deformed and triaxial shapes. One of the examples is the
$A\approx190$ mass region, where both prolate to oblate shape transitions, as well
as triaxial ground-state shapes have been predicted. An extensive analysis
of this region has recently been performed using
non-relativistic Skyrme and Gogny interactions~\cite{RRS.09,RSR.10}. The
self-consistent Hartree-Fock-Bogoliubov model has been used to study the
evolution of the ground-state shapes of Yb, Hf, W, Os and Pt
isotopes. In particular, it has been shown that the isotopic chains with
larger $Z$-numbers in this mass region display a tendency toward triaxial shapes.
Here we present the 3D RHB binding energy maps for the sequence of even-A Pt
isotopes with neutron numbers in the interval from $N=108$ to $N=126$, calculated
with the DD-PC1 energy density functional plus the pairing interaction
Eq.~(\ref{pp-force}).

\begin{table}
\centering
\begin{tabular}{lcccccccccc}
\hline \hline
A & $186$ & $188$ & $190$ & $192$ & $194$ & $196$ & $198$ & $200$ & $202$ & $204$ \\ \hline
$\beta$ & $0.30$ & $0.28$  & $0.19$ & $0.18$ & $0.15$ & $0.13$ & $0.12$ & $0.08$ & 0 & 0 \\
$\gamma$ & $0^\circ$ & $9^\circ$ & $34^\circ$ & $34^\circ$ & $34^\circ$ & $31^\circ$ & $33^\circ$
   & $60^0$ & $0^\circ$ & $0^\circ$
   \\ \hline \hline
\end{tabular}
\caption{\label{tab:pt-gs}Calculated values of the $\beta$ and $\gamma$ deformation parameters 
for the absolute minima of the potential-energy surfaces (PES) of
even-A Pt isotopes with $186 \le A \le 204$.}
\end{table}

\begin{figure}
\centering
\begin{tabular}{cc}
\includegraphics[trim=25mm 5mm 25mm 10mm,clip,scale=0.4]{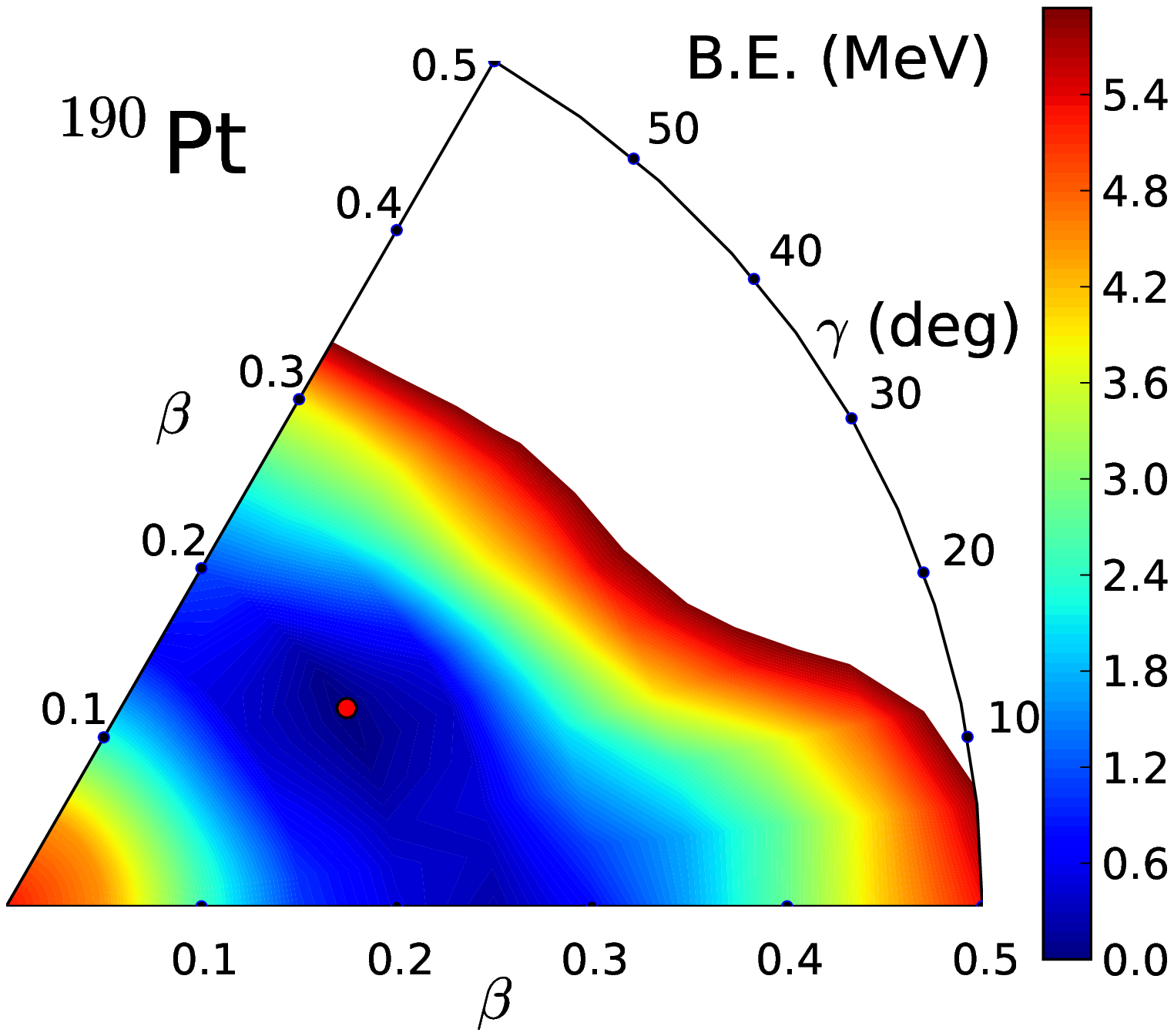}
& \includegraphics[trim=25mm 5mm 25mm 10mm,clip,scale=0.4]{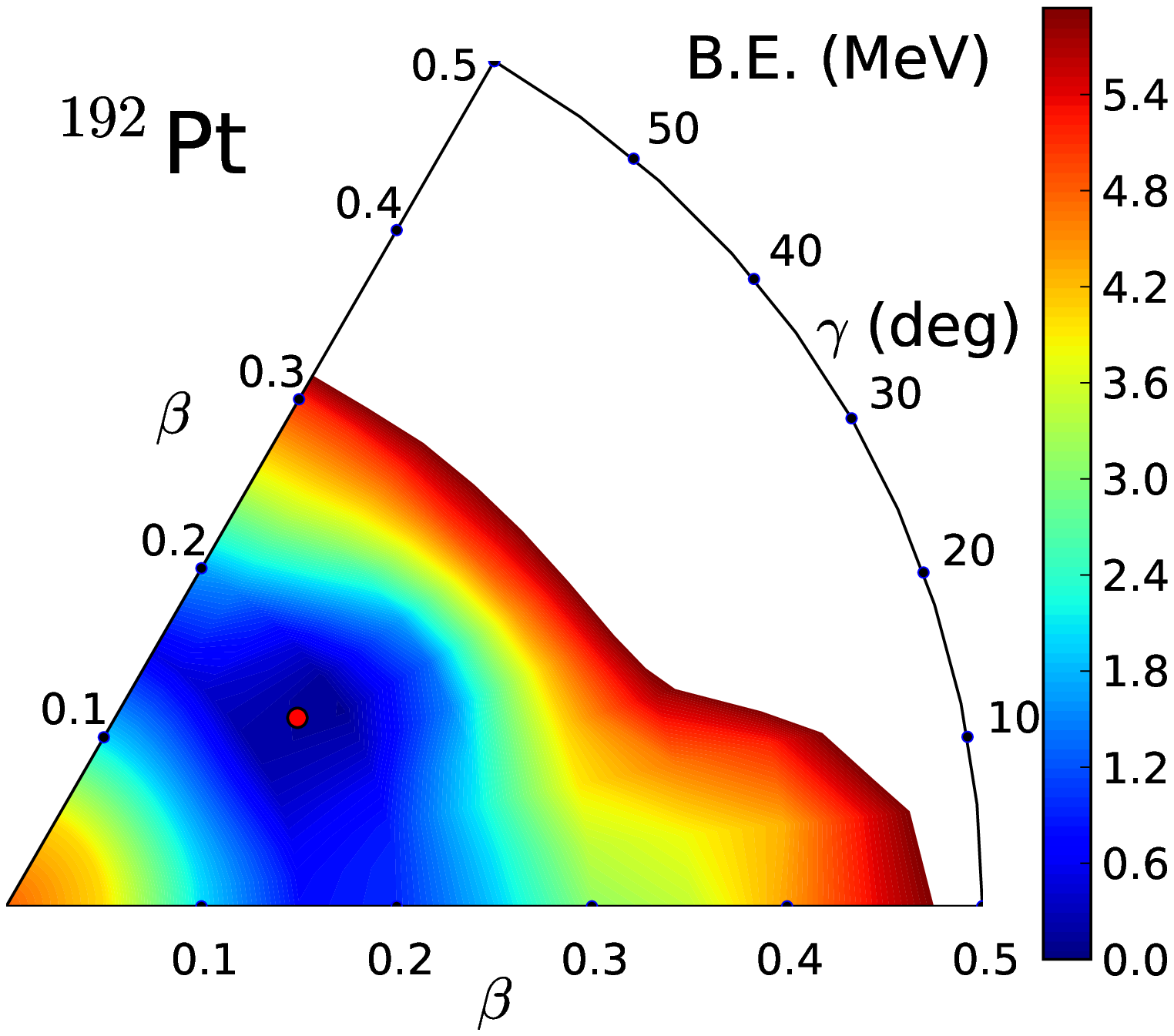} \\
\includegraphics[trim=25mm 5mm 25mm 10mm,clip,scale=0.4]{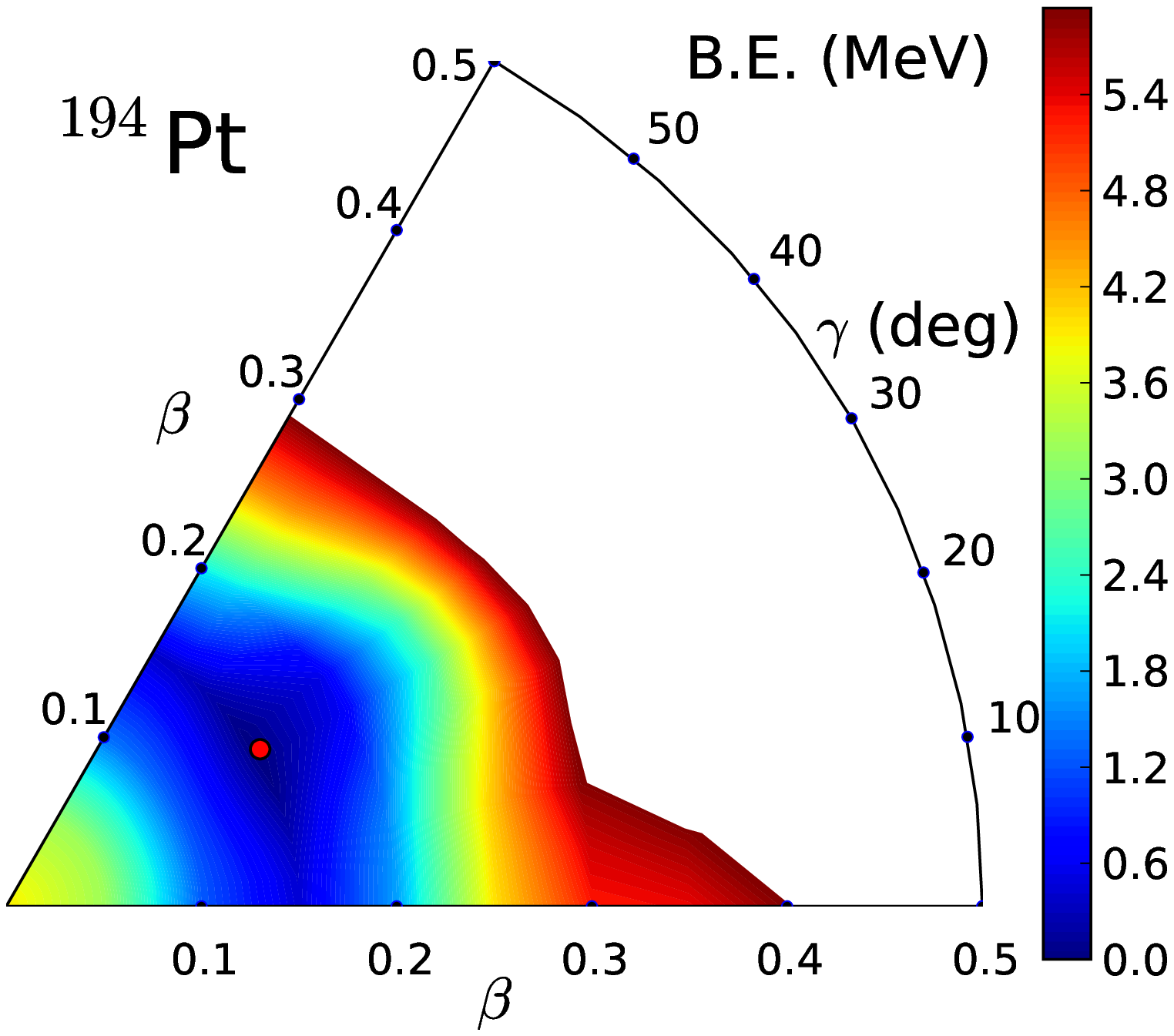}
& \includegraphics[trim=25mm 5mm 25mm 10mm,clip,scale=0.4]{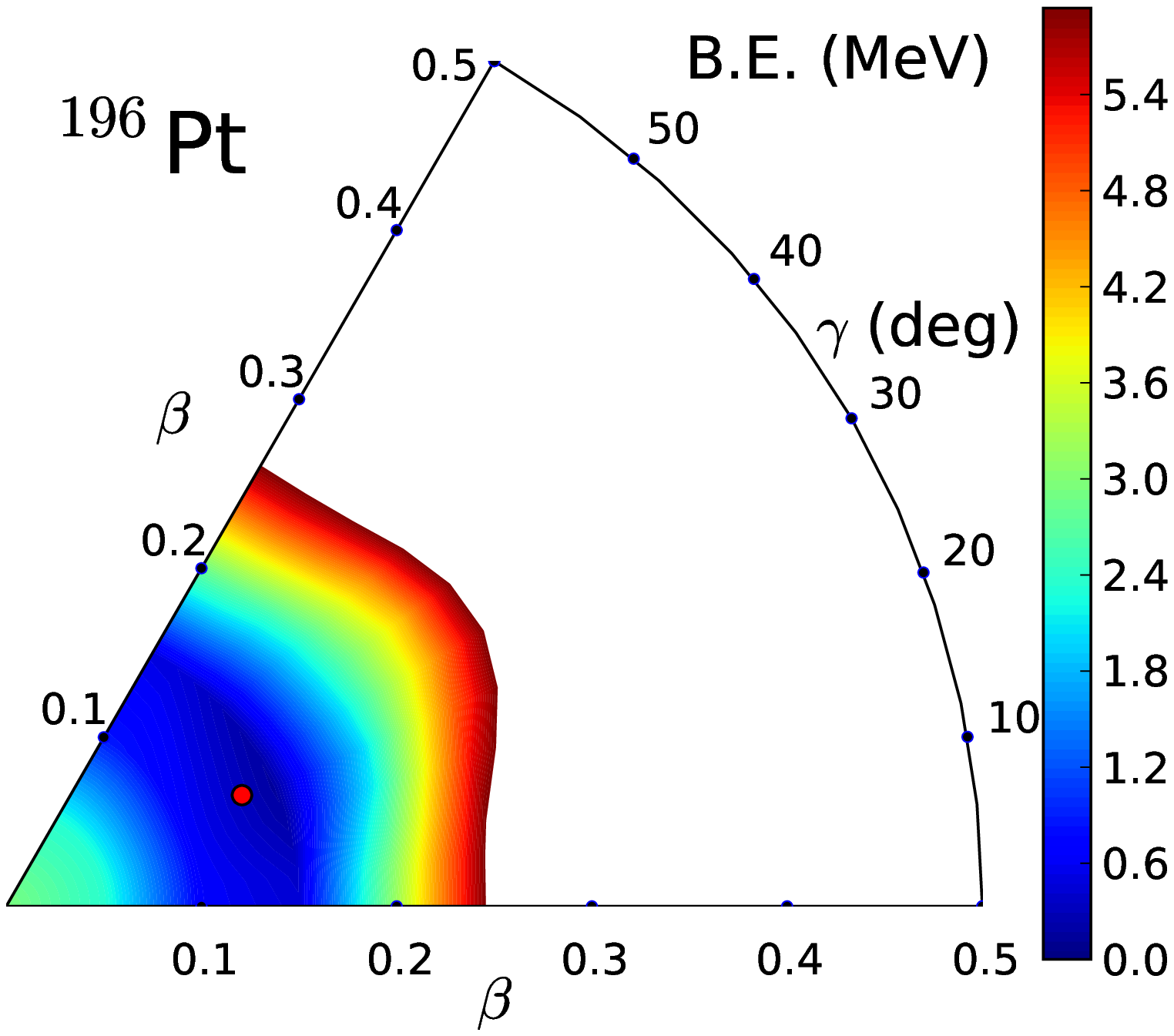} \\
\includegraphics[trim=25mm 5mm 25mm 10mm,clip,scale=0.4]{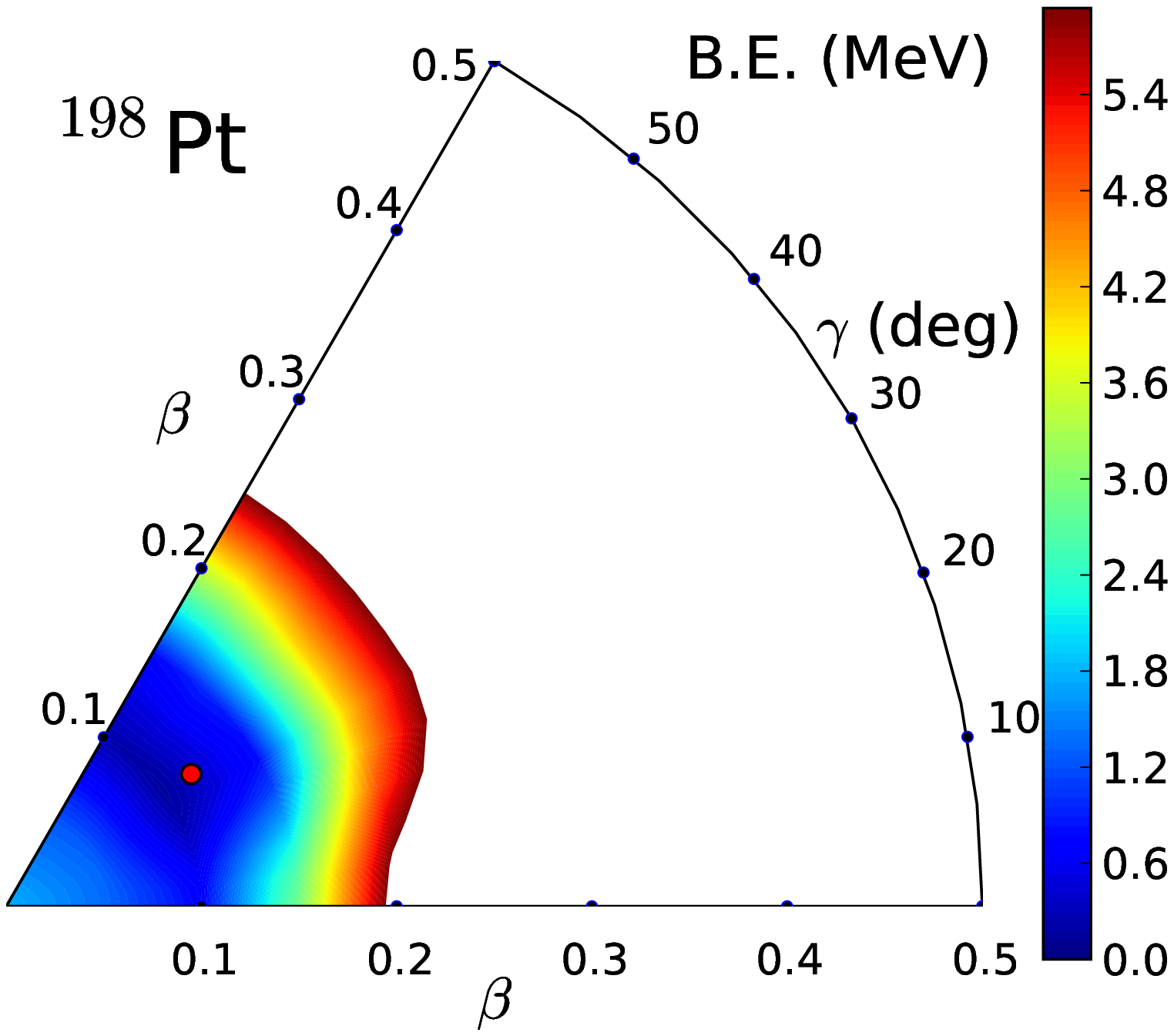}
& \includegraphics[trim=25mm 5mm 25mm 10mm,clip,scale=0.4]{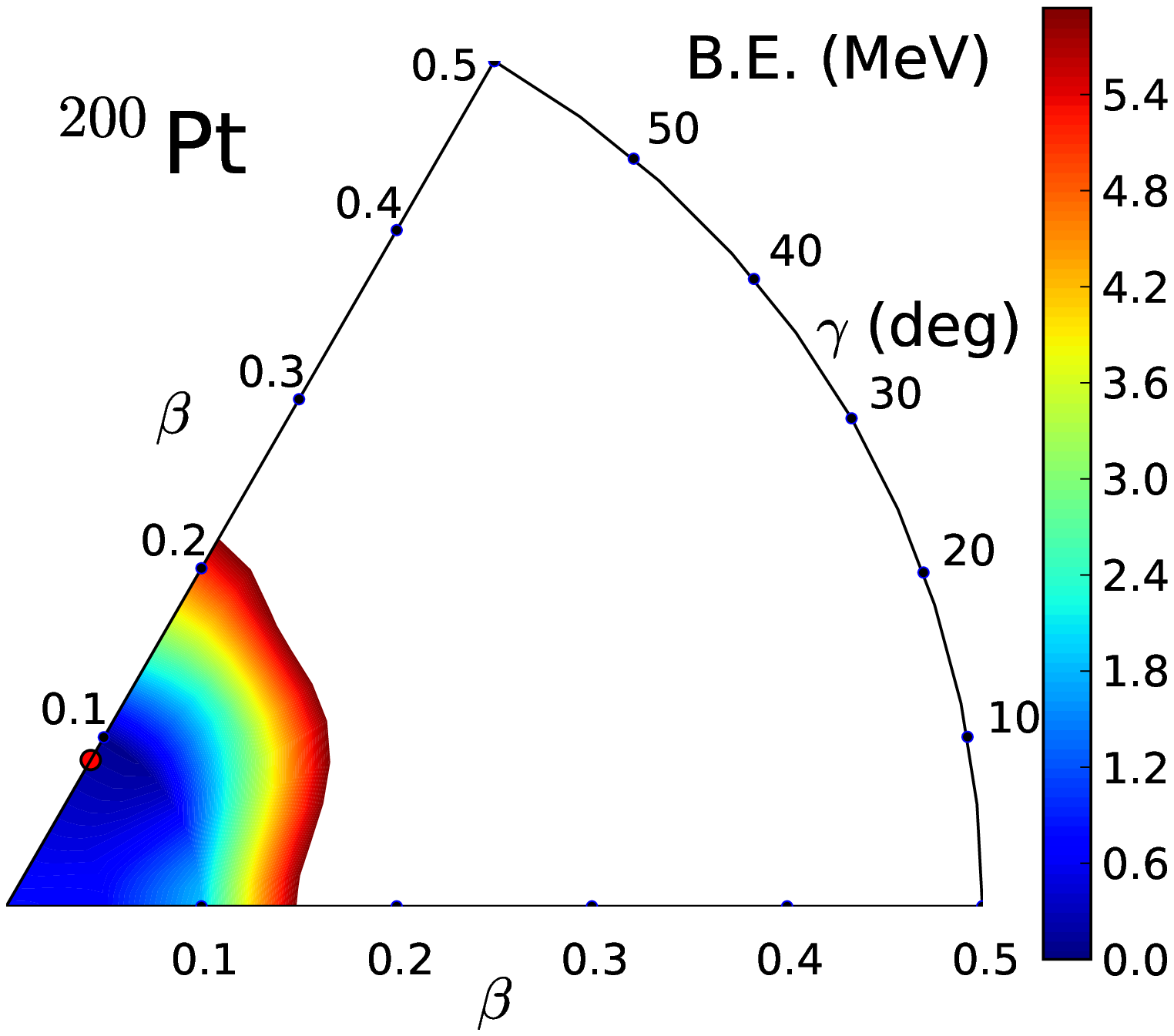} \\
\end{tabular}
\caption{Self-consistent RHB triaxial quadrupole binding-energy maps of the
even-even isotopes $^{190-200}$Pt in the
$\beta-\gamma$ plane ($0\le \gamma \le 60^\circ$). All energies are normalized  with 
respect to the binding energy of the absolute minimum (red dot). }
\label{fig:pes_pt}
\end{figure}
In Tab.~\ref{tab:pt-gs} we list the calculated values of the $\beta$ and $\gamma$
deformation parameters for the absolute minima of  the potential energy surfaces (PES).
One can follow the transition from the prolate deformed $^{186}$Pt,
through the region of triaxially deformed $^{188-198}$Pt isotopes, to the
slightly oblate $^{200}$Pt, and finally the spherical $^{202-204}$Pt isotopes.
The ground-state $\beta$-deformation steadily decreases as
the number of neutrons increases and approaches the closed-shell at $N=126$.
In order to analyze the nature of shape transition in the Pt isotopic chain, in
Fig.~\ref{fig:pes_pt} we display the self-consistent RHB quadrupole
binding energy maps of the even-A $^{190-200}$Pt isotopes in the $\beta-\gamma$ 
plane ($0^{0}\le \gamma \le 60^{0}$). All energies are normalized with respect to the
binding energy of the absolute minimum, and the color code refers to the energy
of each point on the surface relative to the minimum. The PES of $^{190-198}$Pt
are $\gamma$-soft, with shallow minima at  $\gamma\approx30^{0}$.
The nucleus $^{200}$Pt displays a slightly oblate minimum, signaling the
shell-closure at $N=126$.

\begin{figure}
\centering
\includegraphics[scale=0.5]{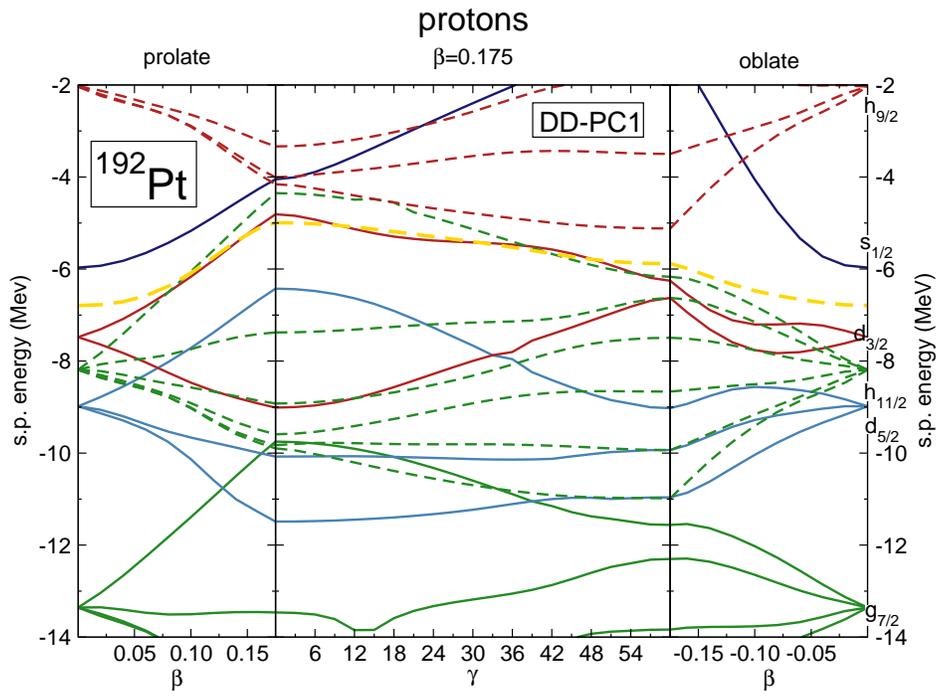}
\caption{\label{Fig:sp-protons-pt192}
Proton canonical single-particle energy levels of $^{192}$Pt. Solid curves
denote levels denote levels with positive parity and short-dashed curves levels with negative parity. 
The long-dashed (yellow) curve corresponds to the Fermi level. The leftmost and the rightmost 
panels display prolate and oblate axially symmetric single-particle levels, respectively. The middle 
panel shows the single-particle levels as functions of $\gamma$ for the fixed value of the axial deformation
$|\beta|=0.18$.}
\end{figure}

\begin{figure}
\centering
\includegraphics[scale=0.475]{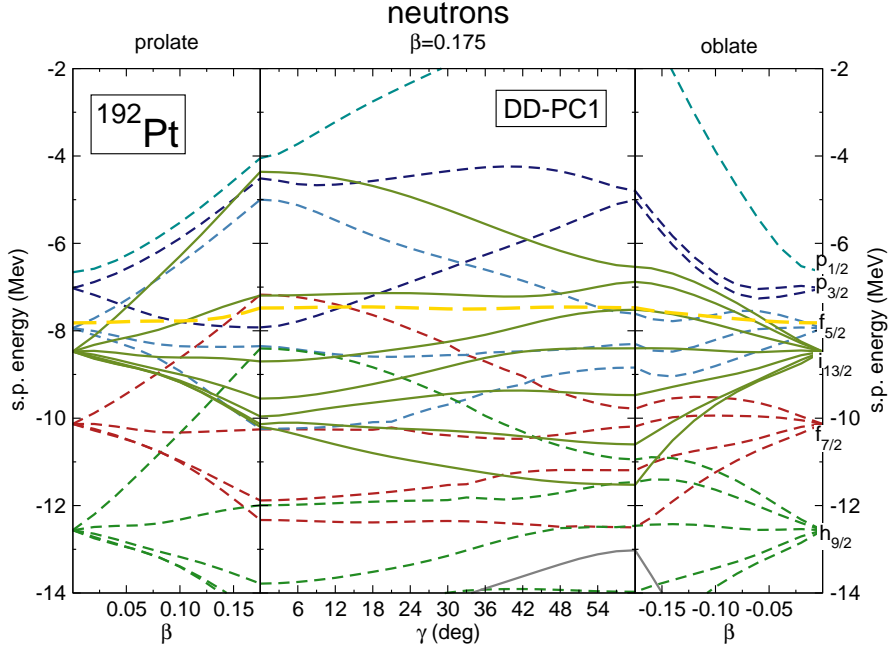}
\caption{\label{Fig:sp-neutrons-pt192}
Same as described in the caption to Fig.~\ref{Fig:sp-protons-pt192} but for neutron single-particle levels.}
\end{figure}

As an illustrative example for the microscopic origin of the triaxial ground-state deformations,
we consider the nucleus $^{192}$Pt.
The formation of deformed minima can be related to the occurrence of gaps or
regions of low single-particle level density around the Fermi surface. In
Figs.~\ref{Fig:sp-protons-pt192} and \ref{Fig:sp-neutrons-pt192} we plot the
proton and neutron single-particle energy levels in the canonical basis for
$^{192}$Pt. Solid curves correspond to levels with positive parity, and
short-dashed curves denote levels with negative parity.
The long-dashed (yellow) curve corresponds to the Fermi level.
The leftmost and the rightmost panels display prolate and oblate
axially-symmetric single-particle levels, respectively, whereas
the middle panel shows the single-particle levels as functions of
$\gamma$ for the fixed value of the axial deformation $|\beta| = 0.18$.
This type of plot has been introduced in Ref.~\cite{Eks.80}, and it enables the
identification of $K$ quantum numbers of triaxial single-particle levels in the
limits of axial symmetry at $\gamma=0^{0}$ and $\gamma=60^{0}$ \cite{RRS.09,RSR.10,CHN.05}.

In Fig.~\ref{Fig:sp-protons-pt192} we notice the occurrence of a gap
between the proton single-particle levels in the vicinity of the Fermi surface
around $\gamma=30^{0}$. The energy gap predominantly results from
the down-sloping of one particular single-particle orbital,
originating from the spherical d$_{5/2}$ shell, as the
deformation parameter $\gamma$ increases from $\gamma=0^{0}$ to $\gamma
=60^{0}$. This result is in agreement with the findings of
Ref.~\cite{RRS.09}. The corresponding neutron single-particle levels, shown
in Fig.~\ref{Fig:sp-neutrons-pt192}, also display a region of low level density around
the Fermi surface at $\gamma \approx 30^{0}$, although the gap is somewhat less
pronounced in comparison to the proton gap.

\begin{figure}
\centering
\includegraphics[scale=0.5]{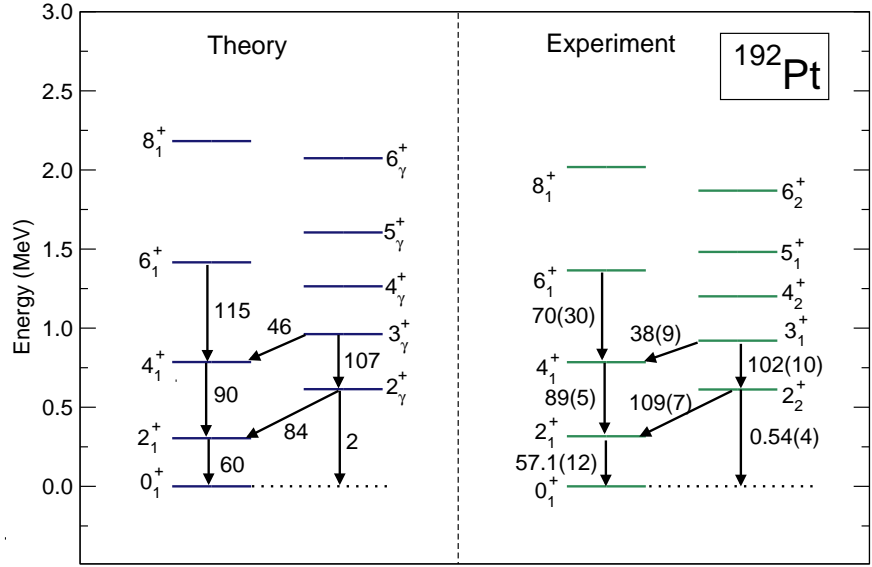}
\caption{\label{Fig:192-spec} The low energy spectrum of $^{192}$Pt calculated with the 
DD-PC1 relativistic density functional (left) compared with the data (right) for
the excitation energies and intraband and interband B(E2) values (in Weisskopf units). }
\end{figure}

\begin{figure}
\centering
\includegraphics[scale=0.5]{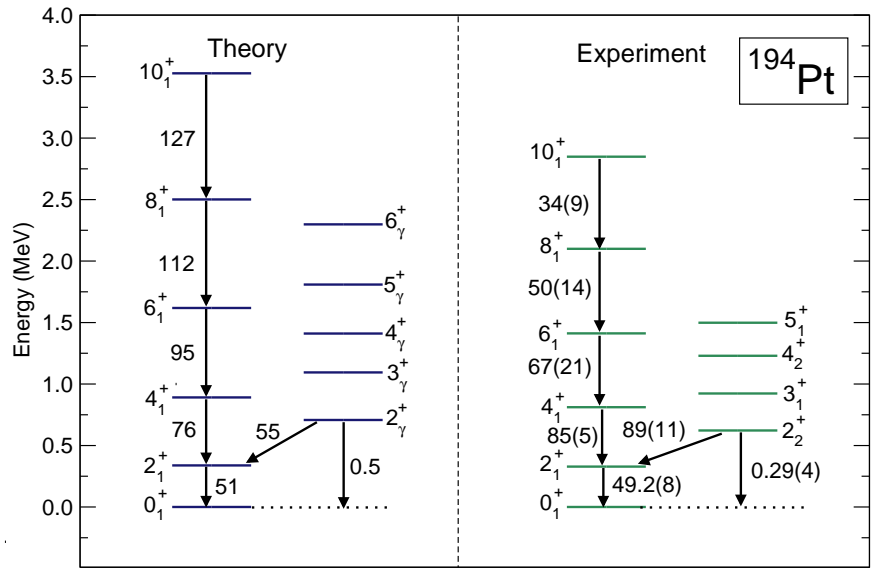}
\caption{\label{Fig:194-spec} Same as described in the caption to
Fig.~\ref{Fig:192-spec} but for the  isotope $^{194}$Pt.}
\end{figure}

The low-energy spectra of $^{192}$Pt and $^{194}$Pt, obtained by
diagonalization of the collective Hamiltonian based on the DD-PC1 energy density
functional plus the pairing interaction Eq.~(\ref{pp-force}), are displayed in
Figs.~\ref{Fig:192-spec} and \ref{Fig:194-spec}. The calculated ground-state bands
and (quasi) $\gamma$-bands are compared with the corresponding sequences of
experimental states. Both the theoretical excitation energies and B(E2) values are
in very good agreement with data. In the case of Pt isotopes it was not necessary to
renormalize the effective moments of inertia, i.e. the spectra shown in
Figs.~\ref{Fig:192-spec} and \ref{Fig:194-spec} do not include any additional
scaling parameter.
As already emphasized, transition probabilities are calculated
in the full configuration space with bare proton charges. In particular, one might
notice the excellent result for the predicted excitation energy of the bandhead
of the $\gamma$-band in both nuclei, as well as the very good agreement
with the experimental B(E2) values for transitions between the $\gamma$-band
and the yrast band. This result indicates that the DD-PC1 potential has the
correct stiffness with respect to the $\gamma$ degree of freedom. A similar
agreement with data is also obtained in the calculation of low-energy spectra
for the other Pt isotopes whose PES are displayed in Fig.~\ref{fig:pes_pt}.


The five-dimensional collective Hamiltonian for quadrupole vibrational and rotational
degrees of freedom, with parameters determined by constrained self-consistent
relativistic mean-field calculations for triaxial shapes, has also been employed in studies
of microscopic signatures of ground-state shape phase transitions in Nd isotopes
around $N=90$ \cite{Li.09a,Li.09b}, and Ba and Xe nuclei in the mass $A \approx 130$
region \cite{Li.10a}. A detailed comparison with available data and with predictions
of the analytical X(5) \cite{FI.01} and E(5) \cite{FI.00} models, has shown that the
microscopic theoretical framework based on relativistic EDFs describes not only
general features of shape transitions, but also singular properties of excitation
spectra and transition rates at the points of first- and second-order quantum shape
phase transitions.

There are possible improvements and extensions of the model that
has been reviewed in this section. For instance, because in most cases
the Inglis-Belyaev formula yields effective moments of inertia that are
lower than empirical values, all the calculated relative
excitation energies had to be scaled with respect
to the experimental energy of the $2_1^+$ states. The moments of
inertia can be improved by including the Thouless-Valatin dynamical
rearrangement contributions. For the rotational degrees of freedom for
which the collective momenta are known, the inertia parameters can be
obtained from the solutions of cranked RHB equations. For the deformation
coordinates $q_0$ and $q_2$ the situation is more complicated,
because the corresponding momentum operators  $\hat{P}_0$ and
$\hat{P}_2$ have to be calculated from the solution of Thouless-Valatin
equations \cite{TV.62} at each deformation point.
Because cranking breaks time-reversal symmetry, in both cases the
inclusion of pairing correlations necessitates calculations in the full
relativistic Hartree-Bogoliubov framework, including time-odd components
of the nucleon self-energies.
\section{Summary and outlook}
\label{Summary}
A wealth of new data from radioactive-beam facilities, the exciting
phenomenology of nuclear astrophysics, and recent theoretical developments
in related fields, have prompted important advances in theoretical nuclear
structure physics during the last decade. The objective of this field is to build
a consistent microscopic framework that will, on the one hand, bridge the gap between
the underlying theory of strong interactions and the phenomenology of finite nuclei
and, on the other, provide a unified description of bulk properties, excitations and reactions
across the entire chart of nuclides.

Even though {\em ab initio} approaches, starting from a microscopic nuclear Hamiltonian
that accurately reproduces scattering and few-body data, have been very successful in the
description of relatively light nuclei up to oxygen isotopes, and large-scale semi-microscopic
shell model calculations are performed for medium-heavy and even some heavy nuclei in the
vicinity of closed shells, at present the only comprehensive approach to nuclear structure
is provided by the framework of energy density functionals. The advantages of using EDFs
in the description of structure phenomena are evident already at the basic level
of implementation -- the self-consistent mean-field method:  an intuitive interpretation of
mean-field results in terms of intrinsic shapes and single-particle states, calculations
are performed in the full model space of occupied states (no distinction between core and
valence nucleons, no need for effective charges), and the universality of EDFs that
enables their applications to {\em all} nuclei throughout the periodic chart. The latter
feature is especially important for extrapolations to regions of exotic short-lived nuclei
far from stability for which few, if any, data are available. For spectroscopic applications,
however, the EDF-based approach must be extended beyond the static mean-field
level and models must be developed that include collective correlations related
to the restoration of broken symmetries in finite nuclei, and take into account
fluctuations of collective variables around mean-field minima.

Relativistic energy density functionals (REDFs) have their origin in the highly successful
relativistic mean-field (RMF) phenomenological models introduced by Walecka and
Serot \cite{SW.86,SW.97}, and later applied and further developed by many groups.
More recently, this framework has been reinterpreted by analogy to relativistic Kohn-Sham
density functional theory. It has been realized that the original meson-exchange forces
used in RMF models present only one of the possible representations of the effective
in-medium inter-nucleon interactions and, moreover, one that does not present any
particular advantage at low energies characteristic for nuclear binding and low-lying
excitations. Functionals have thus been developed that are expressed in terms of
ground-state nucleon four-currents and scalar densities only, with short-distance
correlations and long-range dynamics represented either by higher order powers
of the currents and densities, or encoded in the medium (nucleon density) dependence
of the coupling functions of interaction Lagrangians. The corresponding structure models
have been applied to studies of a variety of phenomena in spherical and deformed
nuclei, extending over the whole mass table and to systems with extreme isospin values.
The illustrative calculations presented in this work have been performed using the
relativistic energy density functional DD-PC1 \cite{NVR.08} for which, starting from
microscopic nucleon self-energies in nuclear matter, the parameters were fine-tuned
in a careful fit to experimental binding energies of 64 axially deformed nuclei in the
mass regions $A\approx 150-180$ and $A\approx 230-250$. For quantitative
calculations in open-shell nuclei it also necessary to consider pairing correlations and,
when used in the relativistic Hartree-Bogoliubov (RHB) framework together with a pairing
force separable in momentum space, the functional DD-PC1 provides an excellent
description of ground-state properties. The corresponding self-consistent QRPA calculations
reproduce the excitation energies of giant multipole resonances.

In this work we have also reviewed the latest extensions of the REDF framework that include
the treatment of collective correlations. By restoring symmetries broken by the static
mean-field and considering fluctuations of collective deformation variables, REDF-based
models have been developed that can be employed in detailed spectroscopic studies,
including predictions for excitation spectra and electromagnetic transitions. For axially
deformed nuclei this approach has been illustrated by GCM configuration mixing calculations
of angular-momentum projected relativistic mean-field wave functions. The GCM excitation
energies and the corresponding B(E2) values for the two lowest bands in $^{154}$Sm
have been discussed in comparison with available data. In an approximation to the full GCM
treatment of the five-dimensional quadrupole dynamics, a collective Bohr Hamiltonian has
been formulated, with deformation-dependent parameters determined by constrained
microscopic self-consistent RHB calculations of triaxial energy surfaces in the
$\beta-\gamma$ plane. The entire map of the energy surface as function of the
quadrupole deformation is obtained by imposing constraints on the
mass quadrupole moments. The quasiparticle wave functions and energies,
generated from constrained self-consistent solutions of the RHB model,
provide the microscopic input for the parameters of the
collective Hamiltonian: the collective potential, the three mass parameters:
$B_{\beta\beta}$, $B_{\beta\gamma}$, $B_{\gamma\gamma}$, and the
three moments of inertia $\mathcal{I}_k$. The implementation of this complex,
REDF-based, microscopic collective model has been exemplified in a study of the
evolution of triaxial shapes in Pt isotopes. The DD-PC1 based
3D RHB calculation provides a simple microscopic interpretation of the
occurrence of triaxial shapes in terms of proton and neutron single-particle levels,
and the calculated ground-state bands and (quasi) $\gamma$-bands are in very
good agreement with the corresponding sequences of experimental states, both for
the excitation energies and B(E2) values.

In the remainder of this section we outline some of the most important challenges for
the framework of (relativistic) density functionals.

An important issue is the development of a series (ladder) of accurate and controlled
approximation for the exchange-correlation terms of the energy density functional,
analogous to the ``Jacob's Ladder" of Coulomb Density Functional Theory \cite{Per.05}.
One possible approach is to develop the exchange-correlation functional from first
principles by incorporating known exact constraints, another is empirical and
optimizes a parametric ansatz by adjusting it to a set of data. In the context of
nuclear density functionals this topic has recently been discussed in the review
of Ref.~\cite{DFP.10}, and a possible non-empirical approach to climbing the
rungs of the ladder  of approximations toward the universal functional has been
indicated.

A related topic is the link between the framework of nuclear EDFs and the
underlying theory of strong interaction -- low-energy QCD, which will include
both nuclear matter and finite nuclei. At low energies characteristic for nuclear
binding, QCD is realized as a theory of pions coupled to nucleons \cite{EHM.09}.
The basic concept of a low-energy effective field theory (EFT) is the separation of
scales: the long-range physics (pion exchange) is treated explicitly, whereas
short-distance interactions, that cannot be resolved at low energy, are replaced by
contact terms. In a non-empirical approach to nuclear NEDFs, the EFT-based
derivation of exchange-correlation functionals in principle allows for error
estimates, and provides a power counting scheme that separates long-
and short-distance dynamics.

Even a non-empirical universal energy density functional will have to be fine-tuned
to data on medium-heavy and heavy nuclei. This is because such a functional
contains a number of low-energy constants that determine the strength of the
leading short-range interactions. These constants cannot be adjusted already
from scattering and few-body data to an accuracy that enables a quantitative
description of structure phenomena in complex nuclear systems. In principle
any complete set of low-energy data, for instance experimental masses, can
be used to fine-tune the EDF. However, when only a small number of nuclei
is considered, satisfactory {\em least-squares} fits can be obtained with
different, in general linearly dependent combinations of parameters. Moreover,
ground states of spherical nuclei that have mostly been used to adjust functionals
or effective interactions, include collective correlations that cannot be absorbed
in a universal functional. Another important issue
concerns functionals that are used in models that go beyond the mean-field
level and include collective correlations. For instance, if rotational energy corrections
and quadrupole fluctuations are treated explicitly by angular momentum projection
and configuration mixing, they should not at the same time implicitly be included in
the functional, i.e. through parameters adjusted to data that already include these
correlations. Therefore, the parameters of such functionals must be adjusted to pseudodata,
obtained by subtracting correlation effects from experimental masses and radii.

In the development of EDF-based structure models that include collective correlations
through symmetry restorations and configuration mixing, relativistic functionals face the
same challenges as their non-relativistic counterpart, i.e. Skyrme-type functionals.
For these models to be able to make spectroscopic predictions in medium-heavy
and heavy nuclei, often characterized by soft potential energy surfaces, it is important to build
accurate and efficient algorithms that perform a complete restoration of symmetries broken
by the static mean field (translational, rotational, particle number), and take into
account fluctuations around the mean-field minima for very general shapes.

An interesting recent development, which goes beyond the scope of the present
review, is the extension of the relativistic (Q)RPA to the quasiparticle time-blocking
approximation \cite{LR.06,LRT.07,LRV.07,LRT.08,LRT.09}. This approach takes
into account effects of particle-vibrational coupling and, therefore, enables a quantitative
analysis of single-particle excitations in odd-mass nuclei and vibrational excitations.
In particular, the dynamics of particle-vibrational coupling leads to an
increase of the level density near the Fermi surface, i.e. to an enhancement of the nucleon
effective mass. The RQRPA extended by the coupling to collective vibrations generates
spectra with a multitude of two-quasiparticle-plus-phonon states, that are important in
the description of damping phenomena characteristic for giant multipole excitations,
as well as in studies of low-energy modes in neutron-rich nuclei.

Finally, interesting results could also be obtained by expanding nuclear energy
density functionals to include non-nucleonic
degrees of freedom. For instance, it has been shown that relativistic density functionals
provide a natural framework for the description of hypernuclear single-particle spectra
based on chiral SU(3) dynamics \cite{FKV.07,FKV.09}.

\bigskip
\noindent
{\bf Acknowledgements}

We would like to thank G. A. Lalazissis, Z. P. Li, Z. Y. Ma, J. Meng, 
L. Pr\'{o}chniak, Y. Tian, and J. M. Yao for their contribution to the 
work reviewed in this article. 
This work was supported in part by the MZOS - project 1191005-1010,
and the DFG cluster of excellence \textquotedblleft Origin and
Structure of the Universe\textquotedblright\ (www.universe-cluster.de).
T. N. acknowledges support by the Croatian Science Foundation.

\newpage

\end{document}